\documentclass[aps,prx,reprint,superscriptaddress,groupedaddress,nofootinbib]{revtex4-2}

\usepackage[english]{babel}

\usepackage{amsmath}
\usepackage{amssymb}
\usepackage{amsfonts}
\usepackage{amsthm}
\usepackage{mathtools}
\usepackage{mathrsfs}
\usepackage{color,verbatim,graphicx}
\usepackage{url}
\usepackage{multirow}
\usepackage[colorlinks=true,citecolor=blue,linkcolor=blue,urlcolor=blue]{hyperref}
\usepackage{cleveref}
\usepackage{diagbox}
\usepackage{booktabs}
\usepackage[detect-weight=true,detect-family=true]{siunitx}
\usepackage{csquotes}
\usepackage{microtype}
\usepackage{multirow}
\usepackage{txfonts}
\usepackage{dsfont}

\newcommand{\ket}[1]{\ensuremath{\left|#1\right\rangle}}
\newcommand{\ketf}[1]{\ensuremath{|#1\rangle}}
\newcommand{\bra}[1]{\ensuremath{\left\langle#1\right|}}
\newcommand{\braket}[2]{\ensuremath{\left\langle #1 \middle| #2 \right\rangle}}
\newcommand{\braketf}[2]{\ensuremath{\langle #1 | #2 \rangle}}
\newcommand{\ketbra}[2]{\ensuremath{\ket{#1}\!\bra{#2}}}
\newcommand{\braopket}[3]{\ensuremath{\left\langle #1 \middle| #2 \middle| #3 \right\rangle}}
\newcommand{\braopketf}[3]{\ensuremath{\langle #1 | #2 | #3 \rangle}}
\newcommand{\proj}[1]{\ketbra{#1}{#1}}

\newcommand{\defeq}{\coloneqq}
\newcommand{\eqndef}{\eqqcolon}

\DeclareMathOperator*{\medotimes}{\text{\raisebox{0.25ex}{\scalebox{0.8}{$\bigotimes$}}}}

\newcommand{\tr}{\operatorname{tr}}

\newcommand{\idop}{\mathds{1}}
\newcommand{\ee}{\mathrm{e}}
\newcommand{\ii}{\mathrm{i}}
\newcommand{\dd}[1]{\mathrm{d}{#1}}
\newcommand{\ddtwo}[1]{\mathrm{d}^2{#1}}
\newcommand{\ddthree}[1]{\mathrm{d}^3{#1}}

\newcommand{\vv}[1]{\boldsymbol{{#1}}}
\newcommand{\vvn}[1]{\hat{\vv{{#1}}}}
\newcommand{\twonorm}[1]{\left\lvert{#1}\right\rvert}
\newcommand{\absval}[1]{\twonorm{#1}}

\newcommand{\conj}[1]{\overline{#1}}

\newcommand{\hc}{\textrm{h.\,c.}}
\newcommand{\nbar}{\bar{n}}
\newcommand{\krondelta}[2]{\mathrm{\delta}_{#1 #2}}

\newcommand{\acronym}[1]{\textsc{\MakeTextLowercase{#1}}}

\newcommand{\na}{\acronym{NA}}

\newcommand{\srplus}{{\textsuperscript{88}}Sr\textsuperscript{+}}

\newcommand{\citeref}[1]{ref.~\cite{#1}}
\newcommand{\figpart}[1]{\emph{#1}}

\begin{document}

\title{Recoil-induced errors and their correction\\in photon-mediated entanglement between atomic qubits}
\author{Jan Apolín}
\altaffiliation[Present address: ]{Institute for Quantum Electronics, Department of Physics,
ETH Zürich, Otto-Stern-Weg 1, 8093 Zürich, Switzerland.}
\author{David P. Nadlinger}
\email{david.nadlinger@physics.ox.ac.uk}
\affiliation{Department of Physics, University of Oxford, Clarendon Laboratory, Parks Road, Oxford OX1 3PU, U.K.}

\begin{abstract}\noindent%
  Photonically-interconnected matter qubit systems have wide-ranging applications across quantum science and technology, with entanglement between distant qubits serving as a universal resource. While state-of-the-art heralded entanglement generation performance thus far has been achieved in trapped atomic systems modelled as stationary emitters, the improvements to fidelities and generation rates demanded by large-scale applications require taking into account their motional degrees of freedom. Here, we derive the effects of atomic motion on spontaneous emission coupled into arbitrary optical modes, and study the implications for commonly-used atom–atom entanglement protocols. We arrive at a coherent physical picture in the form of \enquote{kick operators} associated with each instant in the photonic wavepackets, which also suggests a method to mitigate motional errors by disentangling qubit and motion post-herald. This proposed correction technique removes overheads associated with the thermal motion of atoms, and may greatly increase entanglement rates in long-distance quantum network links by allowing single-photon-based protocols to be used in the high-fidelity regime.
\end{abstract}

\maketitle

\section{Introduction}

Hybrid quantum systems, combining the complementary strengths of matter-based qubits for stationary storage and processing with photonic qubits as flexible interconnects, are of interest for a broad range of applications, including quantum cryptography and communication tasks~\cite{gisinQuantumCryptography2002,Kimble2008}, modular large-scale computing~\cite{ciracDistributedQuantumComputation1999,jiangDistributedQuantumComputation2007,Nickerson2014,Monroe2014}, tests of fundamental physics~\cite{Hensen2015,rosenfeldEventReadyBellTest2017}, and distributed optical interferometry~\cite{khabiboullineOpticalInterferometryQuantum2019}.
Maximally-entangled states between remote qubits serve as a universal resource for such systems; other primitives can be built on top of it, for instance using teleportation protocols~\cite{gottesmanDemonstratingViabilityUniversal1999,olmschenkQuantumTeleportationDistant2009,pfaffUnconditionalQuantumTeleportation2014,mainDistributedQuantumComputing2025}.
Heralded remote entanglement schemes, where success is indicated by the observation of one or more photons emitted from the matter nodes after interference on a beam splitter~\cite{zukowskiEventreadydetectorsBellExperiment1993}, avoid errors due to link loss and have been demonstrated in a number of platforms, including trapped ions~\cite{moehringEntanglementSingleatomQuantum2007}, neutral atoms~\cite{Ritter2012}, defect centres in diamond~\cite{bernienHeraldedEntanglementSolidstate2013,bhaskarExperimentalDemonstrationMemoryenhanced2020}, solid-state quantum dots~\cite{delteilGenerationHeraldedEntanglement2016}, and rare-earth defects in crystals~\cite{ruskucMultiplexedEntanglementMultiemitter2025}.
Trapped-ion demonstrations have reached remote entanglement fidelities of $\SI{97}{\percent}$~\cite{mainDistributedQuantumComputing2025} and rates of $\sim\!\!\SI{200}{\second^{-1}}$~\cite{oreillyFastPhotonMediatedEntanglement2024,stephensonHighRateHighFidelityEntanglement2020}.
While this represents the state of the art in remote entanglement generation to date, both rates and fidelities still fall short of the demands of future large-scale quantum information processing applications~\cite{Monroe2013,coveyQuantumNetworksNeutral2023,rametteFaulttolerantConnectionErrorcorrected2024,liHighRateHighFidelityModular2024}.

\begin{figure}
  \includegraphics{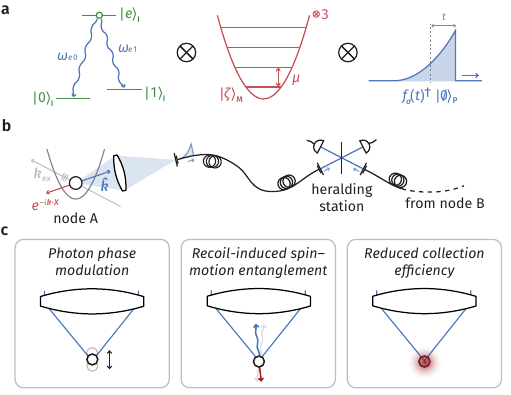}%
  \vspace{-5pt}
  \caption{Spontaneous emission in trapped atoms. \figpart{(a)} We consider atoms decaying from an excited state to one or more stable ground states, involving internal atomic states, centre-of-mass motion, and excitations in the photonic multi-mode vacuum. \figpart{(b)} To generate entanglement between two remote quantum information processing nodes, the trapped atoms are simultaneously excited and the emission is combined at a central heralding station. \figpart{(c)} Atomic motion impacts the remote entanglement generation process through multiple interlinked effects related to recoil during excitation and photon emission; we describe its consequences in a unified framework.}
  \label{fig:scenario}
\end{figure}

Laser-cooled, trapped atoms are well-suited as quantum network interfaces: they are indistinguishable, offer long coherence times as quantum memories, and support high-fidelity logic gates as part of larger-scale devices.
Furthermore, spontaneous emission from short-lived states provides a native optical interface (optical cavities can be employed to tailor the emission process~\cite{Ritter2012,reisererCavitybasedQuantumNetworks2015,gotoFigureMeritSinglephoton2019,schuppInterfaceTrappedIonQubits2021}; here, we focus on free-space configurations).
In previous work on remote entanglement generation, trapped atoms have mostly been treated as ideal, stationary dipole emitters, neglecting their quantum-harmonic motion within the confining potential entirely or treating it in a simplified, semi-classical fashion.
While phase modulation and recoil effects associated with atomic motion had already been recognised as an inherent source of errors in early proposals~\cite{Cabrillo1998}, other, technical limitations to the fidelity observed in demonstration experiments have overshadowed this issue until recently.
As practical realisations are starting to target entangled-state infidelities at or below the percent-level, however, while also aiming to maximise the entanglement generation rate – and thus to minimise time spent on laser cooling –, a detailed understanding of the role of motion becomes essential.
Moreover, even for ground-state-cooled atoms, motional effects fundamentally limit the achievable fidelity. The recoil energy associated with spontaneously emitted photons tends not to be entirely negligible on the harmonic oscillator energy scales; furthermore, as the linewidth of the atomic dipole transitions chosen for efficient photon collection is typically larger than the motional frequency, the motional sidebands are not resolved, hence cannot be rejected using spectral filtering.

In this article, we provide an intuitive framework for describing the effect of atomic motion on photon-mediated entanglement generation. We start by giving a coherent description of spontaneous emission from a trapped emitter into free space (\cref{sec:decay-theory}). We then consider the effect of collection into a single optical mode (\cref{sec:collection}), and show that the overall effect of the emission process is described by a coherent superposition of displacement operators, the rotating-frame phase-space direction of which is entangled with the detection time of the emitted photon. We apply this framework to the generation of remote entanglement using a single herald photon (\cref{sec:single-photon}), and to entanglement swapping using two herald photons generated from multiple decay channels (\cref{sec:two-photon-errors}) or in multiple time-bins (\cref{sec:time-bin-encoding}). We discuss the fidelity loss for an initial thermal state of motion, as well as strategies for its correction by disentangling spin and motion post hoc following a heralding event.

\section{Spin-motion entanglement in~spontaneous~decay}
\label{sec:decay-theory}

\newcommand{\mcol}[1]{#1}
\newcommand{\pcol}[1]{#1}
\newcommand{\icol}[1]{#1}
\newcommand{\adag}[1]{\pcol{a^\dagger_{#1}}}
\newcommand{\bdag}[1]{\mcol{b^\dagger_{#1}}}
\newcommand{\polind}{\sigma}
\newcommand{\modeind}{\pcol{\vv{k}\polind}}
\newcommand{\modeindprime}{\pcol{\vv{k}'\polind'}}
\newcommand{\nmodeind}{\pcol{\vvn{k}\polind}}
\newcommand{\fieldcpl}{\pcol{\mathcal{E}_k}}
\newcommand{\fieldcplcont}{\tilde{\mathcal{E}}_{\pcol{k}}}
\newcommand{\fieldcplcontcentre}[1]{\tilde{\mathcal{E}}_{\icol{\omega_{e#1}} / c}}
\newcommand{\statecpl}[1]{g_{\icol{#1}, \nmodeind}}
\newcommand{\motfreqelem}{\mcol{\mu_j}}
\newcommand{\lambdicke}[1]{\eta_{#1}}
\newcommand{\knormelem}{\pcol{\hat{k}_j}}

Consider a dipole emitter confined in a three-dimensional trapping potential, such as an atomic ion in a radio-frequency trap or an atom trapped in optical tweezers (\cref{fig:scenario}). For simplicity, we will use \enquote{atom} to refer to any dipole emitter throughout the manuscript, and \enquote{spin} for its discrete internal degrees of freedom. Such a system is modelled by the Hamiltonian
\newcommand{\motvec}[1]{\vv{#1}}
\begin{equation}
  \begin{aligned}
    H\ =&\! \underbrace{\sum_{\icol{i}} \hbar\icol{\omega_i \proj{i}}}_{\textrm{internal energy levels}} + \underbrace{\vphantom{\sum_i}\smash{\sum_{\mcol{j}} \hbar \mcol{\motfreqelem \bdag{j} b_j}}}_{\textrm{motion in trap}}
    + \underbrace{\vphantom{\sum_i}\smash{\sum_{\modeind} \hbar c \pcol{\absval{\vv{k}} \, \adag{\modeind} a_{\modeind}}}}_{\textrm{free e.m.~field}}\ \\
    &-\!\! \underbrace{\icol{\vv{d}} \cdot \pcol{\vv{E}}(\mcol{\motvec{X}})}_{\textrm{dipole interaction}}\!\!\!,
  \end{aligned}
\end{equation}
where $i$ labels the internal states of the emitter, $j$ the harmonic oscillator modes describing the confined motion of the particle ($[b_j, \bdag{j'}] = \krondelta{j}{j'}$) making up the total atom position operator~$\motvec{X}$, and we have expanded the electromagnetic field into plane-wave modes with wavevector $\vv{k}$ and polarisation $\polind$ (here in some nominal quantisation volume, though we shall presently take the continuum limit, and $[a_{\modeind}, \adag{\modeindprime}] = \krondelta{\vv{k}}{\vv{k'}}\krondelta{\polind}{\polind'}$).

Here, we are interested in describing the spontaneous decay dynamics from a single excited state $\ket{e}$ to one or more lower-lying levels $\{\ket{i}\}_i$ assumed to be stable (see \cref{fig:scenario}). The relevant components of the electric dipole interactions are then
\begin{equation}
  \vv{d} \cdot \vv{E}(\motvec{X}) = \ii \sum_{i \modeind}\fieldcpl \statecpl{i} \icol{\ketbra{i}{e}} \ee^{-\ii \vv{k} \cdot \motvec{X}} \adag{\modeind} + \hc,
\end{equation}
where $\statecpl{i} = \braopket{i}{\vv{d}}{e} \cdot \vvn{\epsilon}_{\modeind}$ is the matrix element of the electric dipole operator projected onto the normalised electric field vector $\vvn{\epsilon}_{\nmodeind}$ corresponding to the mode $\nmodeind$, and $\fieldcpl$ is a field strength prefactor incorporating the mode density (splitting up $\vv{k}$ into its magnitude $k$ and unit vector $\vvn{k}$).

\newcommand{\dmulti}[2]{\mathcal{D}_{#1}(#2)}
The $\ee^{-\ii \vv{k} \cdot \motvec{X}}$ term, describing the effect of atomic motion on the coupling to the radiation field, can be interpreted in two equivalent ways: manifestly, as an emitter-position-dependent phase of the resulting field, but because position and momentum are conjugate variables, also as a shift in oscillator momentum due to recoil from the emitted photon. There are two salient parameters for the related dynamics: first, the Lamb–Dicke parameter $\lambdicke{j} = \sqrt{E_{\mathrm{recoil}} / (\hbar \motfreqelem)} = \sqrt{\hbar k^2/(2 M \mcol{\mu_j})}$ giving the recoil energy for emitter of mass $M$ in oscillator quanta for each motional mode $j$,%
\footnote{To elucidate the connection to usual concepts in atomic physics, we have defined the Lamb–Dicke parameter $\lambdicke{j}$ for a constant photon wavevector $k \approx \omega_{ei} / c$, which is an excellent approximation as spontaneous emission from atoms is narrow-band ($\Gamma \ll \omega_{ei}$, $\omega_{ei} \defeq \omega_e - \omega_i$), and the lower states $i$ of interest typically do not differ much in energy, $\omega_{ei} \approx \omega_{ei'}$. This choice merely aids exposition in the remote entanglement generation context; the model is straightforwardly generalised to lower levels with significantly different energies by not stating displacements in terms of a common $\eta_j$.}
and secondly, the ratio $\motfreqelem / \Gamma$ of motional frequency to emission linewidth ($\Gamma \propto \sum_{i \modeind} \absval{\fieldcpl\statecpl{i}}^2$) quantifying the amount of oscillator time evolution during the emission process.

\newcommand{\dispj}{D_j}
In detail, consider the Hamiltonian in the interaction picture regarding the free evolution of internal/motional/field states. Assuming the equilibrium position of the emitter to be in the coordinate origin, we can expand $\ee^{\ii \vv{k} \cdot \motvec{X}} = \medotimes_{j} \dispj(\ii \lambdicke{j} \knormelem)$ in terms of displacement operators $\dispj(\beta) = \ee^{\beta \bdag{j} - \conj{\beta}b_j}$ and the projection $\knormelem$ of the normalised direction $\vvn{k}$ onto each motional mode. Defining a multi-mode displacement $\dmulti{\ii \vvn{k}}{t} \defeq \medotimes_{\mcol{j}}\dispj(\ii \ee^{\ii \motfreqelem t} \lambdicke{j} \knormelem)$, we obtain the interaction Hamiltonian in rotating-wave approximation,
\begin{equation}
  H_{\mathrm{int}}(t) \overset{\textrm{RWA}}{\approx} -\ii \sum_{i \modeind}
  \ee^{\ii(kc - \omega_{ei})t} \,
  \fieldcpl \statecpl{i} \icol{\ketbra{i}{e}} \adag{\modeind} \dmulti{-\ii\vvn{k}}{t} + \hc
  \label{eq:hint-rwa}
\end{equation}

\newcommand{\vac}{\emptyset}
Let the atom initially be in the excited state $\ket{\psi_0}_{IMP} = \ket{e}_I\ket{\zeta}_M\ket{\vac}_P$, where $\ket{\zeta}_M$ is an arbitrary state of motion (w.l.o.g.~assumed to be pure for notational convenience) and $\ket{\vac}_P$ is the photonic multimode vacuum. Transitioning to a mode continuum, the evolution under $H_{\mathrm{int}}$ is readily solved in the Weisskopf–Wigner approximation \cite{weisskopfBerechnungNaturlichenLinienbreite1927,milonniQuantumVacuumIntroduction1994,rzazewskiSpontaneousEmissionExtended1992,fedorovSpontaneousEmissionPhoton2005}. At times much longer than $\Gamma^{-1}$ after the initial excitation, where the excitation has been transferred from the atom to a propagating photon wavepacket, the interaction-picture state is approximately independent of time (see \cref{sec:decay-derivation}). The joint state between atom, motion and photon takes a straightforward form after transforming to time-domain plane-wave modes $\tilde{a}^\dagger_{\polind}(t, \vvn{k})$:
\newcommand{\dmatelcentre}[3]{D_{\vec{#1} \vec{#2}}\left(-\alpha_{\icol{\omega_{e1}}/c \vvn{#3}}\right)}
\newcommand{\intsphere}{\int_{\pcol{S^2}}}
\newcommand{\twp}{t}
\begin{equation}
  \begin{gathered}
    \hspace{-90pt} \ket{\psi}_{IMP} \approx \sum_{i } \fieldcplcontcentre{i} \ket{i}_I \int_0^\infty \ee^{-\left(\frac{\Gamma}{2} + \ii \omega_{ei}\right)\, \twp}\\
    \sum_\polind \intsphere \statecpl{i}\, \dmulti{-\ii \vvn{k}}{\twp} \ket{\zeta}_M \tilde{a}^\dagger_{\polind}(\twp, \vvn{k}) \, \ddtwo{\vvn{k}}\, \dd{\twp} \ket{\vac}_P.\hspace{-45pt}
  \end{gathered}
  \label{eq:long-time-state-free-space}
\end{equation}
This is a coherent superposition of the ground states dipole-connected to the excited state $\ket{e}_I$, entangled with photon wavepackets centred on the respective transition frequency, with an exponential envelope given by the excited state lifetime. Each potential photon detection time $\twp$ along the wavepackets is associated with a recoil momentum kick at that instant – in the rotating frame, this is a displacement of the $j$-th mode in the phase space direction $-\ii \ee^{\ii \motfreqelem \twp}$. The displacement amount $\lambdicke{j} \knormelem$ depends on the Lamb–Dicke parameter for the given mode and the projection of the emission direction $\vvn{k}$ on the mode direction.

\section{Collection of emitted photons}
\label{sec:collection}

\newcommand{\thetamax}{\vartheta_{max}}

\begin{figure}
  \includegraphics{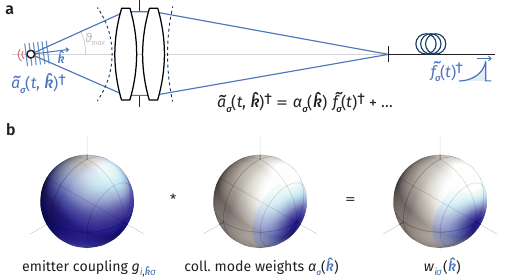}%
  \vspace{-3pt}
  \caption{Collection from a large solid angle into a single mode.
  \figpart{(a)} Typical optical apparatus for remote entanglement generation, where a lens system (numerical aperture $\sin \thetamax$) refocuses emitted photons into a single-mode fibre. In the plane-wave decomposition, each mode $\tilde{a}^\dagger_{\polind}(t, \vvn{k})$ is transformed into a superposition of the selected mode $f^\dagger_{\polind}(t)$ (with coefficient $\alpha_{\polind}(\vvn{k})$, including e.g.~diffraction effects) and some other modes that are subsequently traced out.
  \figpart{(b)} A photon collected into $f_\polind$ hence corresponds to a coherent superposition of emission directions $\vvn{k}$, with the weights $w_{i\polind}(\vvn{k})$ arising as a product of the emission pattern $\statecpl{i}$ and the collection mode profile $\alpha_{\polind}(\vvn{k})$ (shown for dipole emission from an atom in the centre of the sphere, depth of colour indicating amplitudes that happen to be real and positive here).
  }
  \label{fig:collection}
\end{figure}

To access and manipulate the photonic part of the system, a fraction of the dipole radiation emitted into free space is collected using an optical apparatus. The prototypical case is that of a lens objective focussing radiation emitted into a given solid angle quantified by its numerical aperture (\na{}) $\sin\thetamax$ (see \cref{fig:collection}a). In the context of remote entanglement generation, single-mode optical fibres or waveguides are typically employed to transmit emitted photons between emitters and enable efficient interference. This has the effect of selecting only a single transversal mode $F$ with time-domain creation operators $\tilde{f}_\sigma(t)^\dagger$ out of all the photonic modes $P$.

After collection into the single mode $F$ and discarding (tracing out) the other photonic modes, the final state from \cref{eq:long-time-state-free-space} becomes
\begin{multline}
  \label{eq:collected-state}
  \rho_{IM\pcol{F}} \approx \operatorname{proj}\Big(\sum_{i} \ket{i}_I \int_0^\infty \ee^{-\left(\frac{\Gamma}{2} + \ii \omega_{ei}\right)\, t}\\
  \sum_{\polind} K_{i\polind}(t) \ket{\zeta}_M \tilde{f}^\dagger_{\polind}(t)\, \dd{t} \ket{\vac}_F \Big) +
  \rho^{\textrm{rest}}_{IM} \otimes \proj{\vac}_F,
\end{multline}
where $\operatorname{proj}(\ket{\psi})$ denotes a (subnormalised) pure state $\proj{\psi}$ and
\begin{equation}
    K_{i\polind}(t) \defeq \int_{S^2} w_{i\polind}(\vvn{k})\ \dmulti{-\ii \vvn{k}}{t} \,\ddtwo{\vvn{k}},
    \label{eq:all-in-one-op-def}
\end{equation}
while $\rho^{\textrm{rest}}_{IM}$ denotes the atomic state where the photon is not collected, which is not of any further interest here. In \cref{eq:all-in-one-op-def}, $w_{i\polind}(\vvn{k}) \in \mathbb{C}$ gives the contribution of each transversal plane wave mode $\nmodeind$ to the single mode selected%
\footnote{To factor out the transversal modes, i.e.~to write the various quantities as functions of $\vvn{k}$ only, independent of $k = \absval{\vv{k}}$, we have assumed the action of the imaging system to be independent of the photon frequency, which will be true in excellent approximation given the narrow-band nature of atomic spontaneous emission. As with the definition of $\lambdicke{j}$, these assumptions are made for expository purposes only and are not essential to the model.}%
. It arises as a product of the emission pattern given by $\statecpl{i}$, i.e.~the dipole coupling to a particular plane-wave mode, and the contribution $\alpha_{\polind}(\vvn{k})$ of the plane wave $\adag{\polind}(\vv{k})$ to the extracted mode $f_{\polind}^\dagger(k)$ as defined by the transformation describing the action of the imaging system (which we assume to be linear), $\adag{\polind}(\vv{k}) = \alpha_{\polind}(\vvn{k})\, f_{\polind}^\dagger(k) + \ldots$ (see \cref{fig:collection}b). \Cref{eq:collected-state} is easily generalised from pure initial states of motion $\ket{\zeta}_M$ to arbitrary mixed $\rho_M$ using their spectral decomposition.

With this, we have arrived at the central statement of our theoretical framework: the geometrical aspects of spontaneous emission in the combined system of the internal/motional atomic and photonic states are entirely described by the set of \enquote{kick operators} $K_{i\polind}(t)$ (\ref{eq:all-in-one-op-def}) associated to every time offset $t$ within the photonic wavepacket. This description holds for arbitrary initial states of motion, and for arbitrary collection modes, even in the high-\na{} limit.

$K_{i\polind}(t)$ provides a unified description of the phase modulation, recoil, and coupling efficiency modulation effects illustrated in \cref{fig:scenario}b. Note that $K_{i\polind}(t)$ is a superposition of displacement operators; it is not unitary in general. Only in the small-collection-angle (zero-\na{}) limit, where e.g.~for collection along the $z$-axis only, $w_{i\polind}(\vvn{k}) \propto \delta(\vvn{k} - \vvn{z})$, is the overall effect approximated by a simple displacement operator (though with a vanishing prefactor). For non-vanishing collection angles, the non-unitarity of the superposition of displacement operators captures the reduction in collection efficiency from an (e.g.~thermally) smeared-out wavepacket. In general, the motion of the trapped atom will decompose into three (or more) harmonic oscillator modes, and in the high-\na{} case, the projections of the plane-wave components $\vvn{k}$ onto the motional modes will significantly differ, such that $K_{i\polind}(t)$ will leave them in an entangled state.

Nevertheless, the time-resolved detection of a single photon in the output mode at a time offset $t$ projects the joint state into one with known kick operators $K_{i\polind}(t)$ having been applied. Given decay paths to multiple ground states $\ket{i}_I$, or a non-excited state component, this results in spin–motion entanglement. In the following sections, we propose the coherent correction of this, and demonstrate that approximately inverting the kick operators with a spin-dependent displacement is enough to recover remote entanglement fidelity given experimentally relevant parameters. We also make use of the flexibility of $w_{i\polind}(\vvn{k})$ to express arbitrary collection modes to demonstrate that collection into an appropriate standing wave pattern reduces the motional sensitivity (see \cref{sec:two-photon-geometry-errors}).

Note that our discussion has focussed specifically on emission into the mode continuum of free space, from which a single mode is then selected afterwards (without materially changing the mode structure the atom couples to). To treat emission into the mode of a leaky cavity, a non-Hermitian Schrödinger equation can be employed instead~\cite{reisererCavitybasedQuantumNetworks2015,gotoFigureMeritSinglephoton2019}. By deriving appropriate kick operators $K_{i\polind}(t)$ for the propagating output modes, much of the following sections can be generalised to cavity-based systems; we leave this for future work, however.

\section{Single-photon-based entanglement}
\label{sec:single-photon}

\begin{figure}
  \begin{minipage}[b]{0.48\linewidth}
    \includegraphics{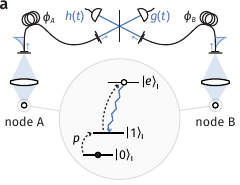}
    \includegraphics{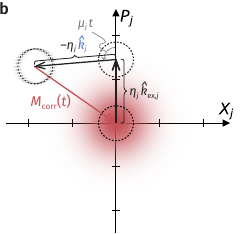}
  \end{minipage}\hfill%
  \begin{minipage}[b]{0.52\linewidth}
    \includegraphics{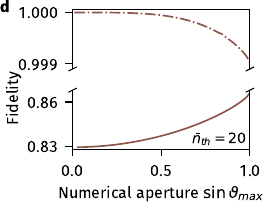}
    \includegraphics{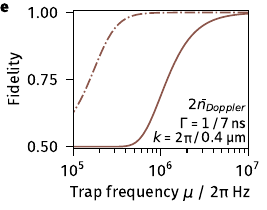}
  \end{minipage}
  \includegraphics{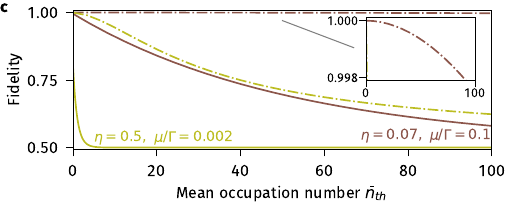}%
  \vspace{-6pt}
  \caption{Single-photon-based remote entanglement.
    \figpart{(a)} Each node starts in a superposition $\sqrt{1-p} \ket{0}_I + \sqrt{p} \ket{1}_I$ and excites only the $\ket{1}_I$ fraction to the emitting state $\ket{e}_I$. For small $p$, the detection of a single photon heralds the creation of an approximate maximally entangled state $\propto \ket{01} \pm \ee^{\ii (\phi_A - \phi_B)} \ket{10}$.
    \figpart{(b)} As depicted in phase space for the $j$-th motional mode, the $\ket{0}_I$ component of the state is not disturbed, while the $\ket{1}_I$ component is displaced both by the excitation laser pulse and the emitted photon. The displacement amounts are given by the product of Lamb–Dicke parameter $\lambdicke{j}$ and the projection of excitation and emission directions $\vvn{k}_{\textrm{ex}}$ and $\vvn{k}$ onto the mode direction; the rotating-frame phase-space direction of the emission kick varies with the detection time as $\ee^{\ii \motfreqelem t}$, where $\motfreqelem$ is the motional mode frequency.
    \figpart{(c)} Heralded Bell state fidelity for isotropic trap frequency/linewidth $\mu / \Gamma$ and Lamb–Dicke parameters $\eta$ representative of ion trap (brown) and optical tweezer (olive) experiments (assuming collection orthogonal to the excitation pulse with \na{} 0.6 into an optimally-sized Gaussian mode).
    The error from spin–motion entanglement (solid lines) can be considerably reduced by overlapping the motional wavepackets with a spin-dependent force $M_{\textrm{corr}}$ after heralding (dash-dotted lines).
    \figpart{(d)} Even outside the zero-\na{} limit, a spin-dependent displacement suffices to heavily suppress motional errors (shown for $\eta = 0.07$, $\mu / \Gamma = 0.1$).
    \figpart{(e)} The spin-dependent correction allows high-fidelity operation at twice the Doppler limit $\hbar \Gamma / (2 k_B)$ for typical trapped-ion cooling transitions, where motional errors are otherwise significant.
  }
  \label{fig:single-photon}
\end{figure}

Given our understanding of the joint atom–photon state after excitation of a single emitter, we now examine the consequences of the motional degrees of freedom on several protocols for heralded photon-mediated entanglement generation, where emission from two remote atoms is brought together and interferes before being detected. Here, we first analyse a class of protocols based on the detection of a single heralding photon~\cite{eichmannYoungsInterferenceExperiment1993,Cabrillo1998}. It is well understood that these protocols are sensitive to the optical phase along the optical link connecting the nodes, and suffer from an intrinsic rate–fidelity tradeoff. Still, such protocols were preferred in several quantum networking demonstrations using solid-state qubits, even if the link phase had to be actively stabilised, as compared to two-photon schemes, their reliance on only a single herald photon leads to significantly improved rates when the overall optical system efficiency is low~\cite{delteilGenerationHeraldedEntanglement2016,stockillPhaseTunedEntangledState2017,humphreysDeterministicDeliveryRemote2018,hermansEntanglingRemoteQubits2023,stolkMetropolitanscaleHeraldedEntanglement2024}.

The errors intrinsic to single-photon schemes, including from an uncontrolled but short-term-stable differential link phase, can be perfectly corrected using a simple single-stage entanglement distillation protocol~\cite{Campbell2008,kalbEntanglementDistillationSolidstate2017}. Two Bell pairs are combined into one with theoretically perfect fidelity, which in the high-loss regime can still yield higher rates than the two-photon protocols discussed later~\cite{Campbell2008}. However, the motion of atomic emitters introduces uncorrelated phase errors that lack the structure to be corrected by this distillation scheme. This issue was already described by Cabrillo et al.~in their original proposal~\cite{Cabrillo1998}, although only for neighbouring atoms in free space; the only experimental realisation to date, using two co-trapped ions, was indeed limited by motional errors~\cite{slodickaAtomAtomEntanglementSinglePhoton2013}. Here, we show how to describe the motional effects for the technologically relevant case of single-mode collection, and that they can be compensated after the fact.

The single-photon scheme is based on emitters with two stable internal states, $\ket{0}_I$ and $\ket{1}_I$ (see \cref{fig:single-photon}a). $\ket{0}_I$ does not participate in the decay process, and only a small fraction $p \in [0, 1]$ is coherently transferred to the short-lived excited state $\ket{e}_I$ (for instance, by preparing a superposition $\sqrt{1-p} \ket{0}_I + \sqrt{p} \ket{1}_I$ and only exciting $\ket{1}_I$ to $\ket{e}_I$). Immediately after excitation (assuming again w.l.o.g.~a pure motional state $\ket{\zeta}_M$ for notational clarity), the joint atom–photon state is
\newcommand{\laserkick}{L}
\newcommand{\kex}{\vvn{k}_{\mathrm{ex}}}
\newcommand{\kexelem}{\hat{k}_{\mathrm{ex},j}}
\newcommand{\kmeanelem}{k_{\mathrm{avg},j}}
\begin{equation}
  \ket{\psi}_{\icol{I}\mcol{M}\pcol{F}} = \sqrt{1 - p} \icol{\ket{0}_I} \mcol{\ket{\zeta}_M} \pcol{\ket{\vac}_F} + \sqrt{p} \icol{\ket{e}_I} \mcol{\laserkick \ket{\zeta}_M} \pcol{\ket{\vac}_F}
\end{equation}
where $L = \dmulti{\ii \vvn{k}_{ex}}{0}$ is the momentum kick of the excitation laser pulse%
\footnote{Here, we assume impulsive excitation to $\ket{e}$ from an approximate plane wave field, as describes well a pulse from a mode-locked laser that is not exceedingly strongly focussed, as often employed in high-rate, high-fidelity remote entanglement experiments.
Given that in the trapped ion/atom context, the linewidth of the dipole transition used to quickly extract photons usually significantly exceeds the motional frequencies ($\Gamma \gg \motfreqelem$ for each mode $j$), high-fidelity excitation to $\ket{e}$ implies this impulsive regime, though the model could be adapted to different excitation schemes by an appropriate choice of $L$.%
}%
~\cite{mizrahiUltrafastSpinMotionEntanglement2013} incident on the atom from direction $\kex$.
After a duration long compared to $\Gamma^{-1}$ has elapsed, the state reads
\begin{multline}
  \label{eq:single-photon-wavepacket}
  \rho_{\icol{I}\mcol{M}\pcol{F}} = \operatorname{proj}\Big( \!\sqrt{1 - p} \icol{\ket{0}}_I \mcol{\ket{\zeta}}_M \pcol{\ket{\vac}}_F + \sqrt{p} \icol{\ket{1}}_I \int_0^\infty \ee^{-\left(\frac{\Gamma}{2} + \ii \icol{\omega_{e1}}\right)\, t} \\
  \mcol{K}(t) \mcol{\ket{\zeta}}_M \pcol{\tilde{f}^\dagger}(t)\, \dd{t} \pcol{\ket{\vac}}_F \!\Big) + p \icol{\proj{1}}_I \mcol{\ldots} \pcol{\proj{\vac}}_F
\end{multline}
with a combined kick operator $\mcol{K}(t) \defeq \mcol{K_{1\sigma}}(t)\, \mcol{\laserkick}$ including both the laser excitation and decay, and where the dots elide the exact form of the component where the atom was excited but the emitted photon was not collected.

The emission from two nodes $A$ and $B$ is then combined on a central $50:50$ beamsplitter (\cref{fig:single-photon}a), and the detection of a single photon in either output mode heralds the creation of a remotely entangled state.
Concretely, in the absence of motional effects, a click for small excitation probabilities $p$ heralds the creation of a maximally entangled state $\approx \ket{01} \pm \ee^{\ii (\phi_A - \phi_B)} \ket{10}$, with the sign depending on the detector, and $\phi_A, \phi_B$ corresponding to the overall optical path length from nodes $A$ and $B$ to the beamsplitter. For finite $p$, double excitation leads to an intrinsic error term $\approx p \ket{11}$.

To include motional effects, consider two wavepackets of the form (\ref{eq:single-photon-wavepacket}), and combine the output modes $\tilde{f}_A(t)$ and $\tilde{f}_B(t)$ on a beam splitter, transforming them as $\tilde{f}_A^\dagger(t) \mapsto (\tilde{g}^\dagger(t) + \tilde{h}^\dagger(t)) / \sqrt{2}$ and $\tilde{f}_B^\dagger(t) \mapsto (\tilde{g}^\dagger(t) - \tilde{h}^\dagger(t) ) / \sqrt{2}$. The success condition for the protocol is a single heralding detector click in either $\tilde{g}(t)$ or $\tilde{h}(t)$, but not both, at any time $t$ (relative to the start of the photon wavepacket following initial excitation, retarded by the optical path from atoms to the detectors, which must be matched between the nodes).
Our model readily yields analytical solutions for the resulting state in full generality (see \cref{sec:single-photon-post-herald-derivation}). For illustrative purposes, we here consider the case where the parameters of both nodes are identical. Assume the state of either node at the beginning of the protocol is $\rho_{IMF} = \proj{0}_I \otimes \rho_M \otimes \proj{\vac}_F$ for some general motional state $\rho_M$, and consider the post-herald state for an infinite heralding window. Tracing out the motional degree of freedom at the end of the protocol, we obtain the following compact expression for the fidelity of the atom–atom state to the closest Bell state%
\footnote{The approximate equality in \cref{eq:single-photon-fidelity} is due to a minor approximation made which holds as long as the overall click efficiency of the protocol is not too high. If the collection efficiency is high, and the photon wavepackets are not pure and identical, sometimes an otherwise-forbidden click in both detectors can be observed. This can be rejected by the apparatus, and hence slightly lowers the influence of the erroneous double-excitation component (see \cref{sec:single-photon-post-herald-derivation}). This slight upward correction of the fidelity is immaterial for practical parameters.}%
:
\begin{equation}
  \label{eq:single-photon-fidelity}
  F \overset{\textrm{eff.}\ll 1}{\approx} (1 - p)\ \frac{1 + C}{2},\
  C \defeq \frac{\int_{0}^{\infty} \ee^{-\Gamma t} \absval{\tr(\mcol{K}(t)\, \mcol{\rho_M})}^2 \dd{t}
  }
  {\int_{0}^{\infty} \ee^{-\Gamma t} \tr(\mcol{K}(t)^\dagger \mcol{K}(t)\, \mcol{\rho_M})\, \dd{t}},
\end{equation}
where the denominator of C, appropriately normalised, gives the collection efficiency.
Using \cref{eq:single-photon-fidelity}, we can easily obtain the fidelity of the generated states for arbitrary motional states and collection geometries through the expectation values of the kick operator. In general, the spatial integral in \cref{eq:all-in-one-op-def} can be evaluated numerically (see \cref{sec:gaussian-mode-calculations}).

In the zero-\na{} limit, where photons are collected only exactly along one axis (and correspondingly, the success probability is infinitesimally small), $K(t)$ is simply a concatenation of two displacements from laser excitation and recoil, so \cref{eq:single-photon-fidelity} is easily evaluated via the characteristic function of the components of $\rho_M$~\cite{barnettMethodsTheoreticalQuantum2005}. Let the excitation laser propagate at angle $\chi$ to the collection axis, and, to obtain a compact expression, let the atom be confined in an isotropic potential with resulting uniform motional frequency $\mu$. For an initial thermal state with mean occupation number $\nbar_{th}$ in each motional mode, we obtain
\begin{equation}
  \label{eq:single-photon-fidelity-lowna}
  C \approx \Gamma \int_{0}^{\infty} \ee^{-\Gamma t}\, \ee^{-2 \eta^2 (1 - \cos(\pcol{\chi}) \cos(\mcol{\mu} t)) (1 + 2 \mcol{\nbar_{th}})}\, \dd{t}.
\end{equation}
For this special case of zero \na{}, our results thus coincide with those derived using a different approach already as part of the original proposal by Cabrillo et al.~for direct interference on a (then vanishingly small) free-space detector~\cite{Cabrillo1998}.

We can understand this as spin–motion entanglement caused by the fact that the kick operator only acts on the emitting $\ket{1}_I$ component, as illustrated in \cref{fig:single-photon}b. As evident from \cref{eq:single-photon-fidelity-lowna}, an excitation pulse propagating in the collection direction, $\chi = 0$, mitigates this by causing the momentum kicks/phase factors to cancel at least for small $t$, but this is often impractical as it amounts to pointing the excitation laser into the collection optics. The fidelity curves in \cref{fig:single-photon}c–e are given for the middling case of orthogonal excitation and collection, $\chi = \pi / 2$.

The impact of this residual entanglement could be reduced by engineering the motional state, e.g.~through ground-state cooling or squeezing. But our analysis suggests a different strategy altogether: after detection of the herald photon at time $t$, the kick operator $K(t)$ applied to the atom in state $\ket{1}_I$ is known. By un-doing it with an appropriate spin-dependent operation in each node, the unwanted spin–motion entanglement can be reduced. This might not be possible in the general high-\na{} case, as the kick operators are complicated superpositions  of multi-mode displacements. Applying just a spin-displacement to each mode, however, is certainly feasible (spin-dependent forces generated using narrow-linewidth lasers or stimulated Raman transitions are part of the standard trapped-atom toolbox~\cite{Wineland1998}). After receiving the timestamp $t$ of the photon detection from the heralding station, each node can apply
\begin{equation}
  \label{eq:single-photon-correction}
  \begin{gathered}
    U_{\mathrm{corr}}(t) = \proj{0}_I \otimes \idop_M + \proj{1}_I \otimes M_{\mathrm{corr}}(t)\textrm{, where}\\
    M_{\mathrm{corr}}(t) = \medotimes\nolimits_{\mcol{j}} \ee^{\ii\, \eta_j^2\, \kexelem\, \kmeanelem \sin(\motfreqelem t)}\, \dispj\!\left(-\ii \,\lambdicke{j}\, (\kexelem - \ee^{\ii \motfreqelem t} \kmeanelem)\right),
  \end{gathered}
\end{equation}
where $\kmeanelem$ is chosen to compensate the emission kick to maximise the fidelity (so outside the zero-\na{} limit a compromise \enquote{average} displacement). $U_{\mathrm{corr}}(t)$ effectively modifies the kick operator $K(t)$ to $M_{\mathrm{corr}}(t) K(t)$. As shown in \cref{fig:single-photon}c–e, for parameters typical of trapped-ion experiments such a simple displacement correction can virtually eliminate motional errors, even at Doppler-cooling temperatures.

Thus, our results demonstrate that single-photon based schemes can feasibly be adapted to achieve low error rates even without ground state cooling, and as such enable a significant increase in entanglement generation rate in the loss-dominated scenarios of interest for long-distance quantum networking.

\section{Two-photon-based entanglement (following single excitation)}
\label{sec:two-photon-errors}

\begin{figure}
  \begin{minipage}[b]{0.4\linewidth}
    \includegraphics{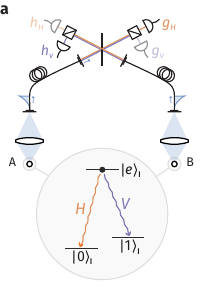}
    \includegraphics{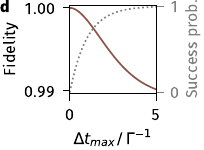}
  \end{minipage}\hfill%
  \begin{minipage}[b]{0.6\linewidth}
    \includegraphics{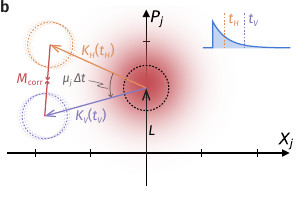}
    \includegraphics{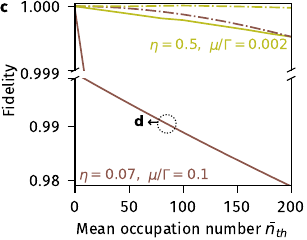}
  \end{minipage}
  \caption{Two-photon-based remote entanglement.
    \figpart{(a)} Both parties excite their atom into the emitting state $\ket{e}_I$; the two decay channels to $\ket{0}_I$ and $\ket{1}_I$ are associated with orthogonal photon modes (here labelled $H$ and $V$). A coincident detection of a photon from both modes in the photonic Bell state measurement setup heralds the creation of a maximally entangled atom–atom state $(\ket{01} \pm \ket{10})\ / \sqrt{2}$.
    \figpart{(b)} As illustrated in the phase space of the $j$-th motional mode of one of the nodes, a difference $\Delta t \defeq t_V - t_H \neq 0$ in detection times for the photons from either decay channel leads to spin–motion entanglement.
    \figpart{(c)} Reduction in fidelity assuming a symmetric trap with the same ratio of trap frequency to linewidth $\mu / \Gamma$ and Lamb–Dicke parameter $\eta$ for all directions, for two sets of parameters characteristic of trapped-ion (brown) and optical tweezer (olive) experiments, given an initial thermal state (excitation pulse orthogonal to the collection direction; collection with numerical aperture 0.6 into an optimally-sized Gaussian fibre mode). Solid lines show the fidelity when ignoring atomic motion, dash-dotted lines the improvement through applying a spin-dependent correction $M_{\textrm{corr}}(t_H, t_V)$ to disentangle spin and motion.
    \figpart{(d)} If only heralds with $\absval{\Delta t} < \Delta t_{max}$ are accepted, the error can also be reduced, but at a significant cost in entanglement generation rate (shown for $\eta = 0.07$, $\mu / \Gamma = 0.1$, $\nbar_{th} = 85$).
  }
  \label{fig:two-photon-timing}
\end{figure}

We now turn our attention towards remote entanglement protocols based on a single (strong) excitation to $\ket{e}_I$ and subsequent decay to the stable states $\ket{0}_I$ and $\ket{1}_I$ along two possible channels with orthogonal photonic modes. Protocols using time-bin encoding are closely related, but involve two separate excitation events; we shall discuss them in~\cref{sec:time-bin-encoding}.

In contrast to the single-photon protocol discussed in \cref{sec:single-photon}, two-photon protocols~\cite{Simon2003,duanEfficientEngineeringMultiatom2003} are not based on weak excitation and can thus work in the high-efficiency regime without an inherent rate/fidelity tradeoff. An (approximately) maximally entangled atom–photon state is generated at each node, and the entanglement is subsequently swapped onto the remote atoms by a Bell-basis measurement on the photons~\cite{zukowskiEventreadydetectorsBellExperiment1993}.
Typically, atomic structure and collection geometry are chosen such that the decay channels are mapped to orthogonal polarisations in the single-mode fibre linking the nodes~\cite{hofmannHeraldedEntanglementWidely2012,nollekeEfficientTeleportationRemote2013,Hucul2014,stephensonHighRateHighFidelityEntanglement2020,krutyanskiyEntanglementTrappedIonQubits2023}, although e.g.~a frequency encoding has also been used~\cite{moehringEntanglementSingleatomQuantum2007}.
The robust nature of two-photon-based protocols, when combined with the well-controlled nature of atomic emitters, has allowed experiments to reach the highest fidelities to date~\cite{mainDistributedQuantumComputing2025}
and, among optical remote entanglement experiments reaching Bell-state fidelities $>\SI{70}{\percent}$, also the highest rates~\cite{stephensonHighRateHighFidelityEntanglement2020,oreillyFastPhotonMediatedEntanglement2024}.

A key feature aiding the robustness of two-photon protocols is that only the relative phase of the wavepackets corresponding to the two decay channels influences of the photonic wavepackets influences the entangled state phase~\cite{Simon2003}; as long as the photons remain temporally overlapped on the Bell-state analyser, slow changes to the absolute optical phase between the nodes can be tolerated. Because of this, such protocols are also relatively less sensitive to motional effects when the emission channels are collected along the same path. However, as other error sources in practical realisations are reduced, the effect of atomic motion is no longer negligible. Here, we derive analytical expressions for the post-herald states and their fidelities in the full picture including motional effects, and provide an intuitive explanation of the error mechanisms in the spin–motion-entanglement picture.

Concretely, consider an atom that has just been excited to the state $\ket{e}_I L\ket{\zeta}_M \ket{\vac}_P$ and subsequently decays along two possible paths to the stable qubit states $\ket{0}_I$ and $\ket{1}_I$ (see \cref{fig:two-photon-timing}a). To aid clarity of exposition, we shall restrict the model such that the two decay channels give rise to excitations in fully orthogonal collection modes $H$ and $V$ (where the notation hints at the common case of emission from two orthogonal dipoles mapping to orthogonal linear polarisations in the single-mode fibre~\cite{Kim2011,stephensonHighRateHighFidelityEntanglement2020}; see \cref{sec:longitudinal-polarisation-symmetry-loss}). The post-decay atom–photon state is then
\begin{equation}
  \label{eq:two-photon-wavepacket}
  \begin{aligned}
    \rho_{\icol{I}\mcol{M}\pcol{F}} = \operatorname{proj}\Big(&\icol{\ket{0}} \int_0^\infty \ee^{-\left(\frac{\Gamma}{2} + \ii \icol{\omega_{e0}}\right)\, t} \mcol{K_H}(t) \mcol{\ket{\zeta}} \pcol{\tilde{f}_H^\dagger}(t)\, \dd{t} \pcol{\ket{\vac}} +\\
    &\icol{\ket{1}} \int_0^\infty \ee^{-\left(\frac{\Gamma}{2} + \ii \icol{\omega_{e1}}\right)\, t} \mcol{K_V}(t) \mcol{\ket{\zeta}} \pcol{\tilde{f}_V^\dagger}(t)\, \dd{t} \pcol{\ket{\vac}}
    \!\Big) \\
    &\hspace{-12pt} + \rho_{IM, \textrm{dark}} \otimes \pcol{\proj{\vac}},
  \end{aligned}
\end{equation}
where $\rho_{IM, \textrm{dark}}$ describes the atomic state for the case where no photon was collected, which is not of interest here due to the later conditioning on a two-photon herald (unless imperfect detectors with dark counts are considered). As in \cref{sec:single-photon}, we have absorbed the momentum kick $L$ from the excitation laser into $K_i(t)$ (which here, however, applies to all components of the superposition).

As illustrated in \cref{fig:two-photon-timing}a, the emission from two nodes $A$ and $B$ is combined in a photonic Bell-state measurement apparatus, consisting e.g.~of a $50:50$ beamsplitter followed by two polarising beamsplitters for to implement polarisation-sensitive single-photon detection. Without considering atomic motion, a two-photon coincidence on two detectors of orthogonal polarisation projects the input state $(\ket{0H} + \ket{1V})^{\otimes 2} / 2$ into a maximally entangled atom–atom state $(\ket{01} \pm \ket{10})\ / \sqrt{2}$ (with the sign depending on the detector combination).

To obtain the state in the full picture including atomic motion, consider two copies of the wavepacket from \cref{eq:two-photon-wavepacket} incident on the Bell-state analyser apparatus (with some arbitrary motional states $\rho_{M^A}$ and $\rho_{M^B}$ in each node). An ideal beamsplitter network maps $\tilde{f}_{Ai}^\dagger(t) \mapsto (\tilde{g}_i^\dagger(t) + \tilde{h}_i^\dagger(t)) / \sqrt{2}$ and $\tilde{f}_{Bi}^\dagger(t) \mapsto (\tilde{g}_{i}^\dagger(t) - \tilde{h}_{i}^\dagger(t) ) / \sqrt{2}$ for either polarisation $i \in \{H, V\}$.
A herald event then consists of two-photon coincidence, where after simultaneous excitation, a photon is detected in both one of $\{g_H, h_H\}$ at time $t_H$ and one of $\{g_V, h_V\}$ at time $t_V$.

The derivation of the full post-herald state is given in \cref{sec:two-photon-post-herald-derivation}. Here, assume again for compactness that the properties of both nodes are the same, and all herald events up to $t \rightarrow \infty$ are accepted. The fidelity of the resulting state to the closest maximally entangled state is then given by
\newcommand{\hcol}[1]{#1}
\newcommand{\vcol}[1]{#1}
\begin{equation}
  \label{eq:two-photon-symmetric-fidelity}
  \begin{gathered}
    F = \frac{1 + C}{2},\textrm{ where}\\
    C \defeq \frac{\int_{0}^{\infty}\int_{0}^{\infty} \ee^{-\Gamma (t_{\!\hcol{H}} + t_{\hspace{-0.4pt}\vcol{V}})} \absval{\tr\!\left(\mcol{K}_{\hspace{-0.4pt}\vcol{V}}(t_{\hspace{-0.4pt}\vcol{V}})^\dagger \mcol{K}_{\!\hcol{H}}(t_{\!\hcol{H}})\, \mcol{\rho_M}\right)}^2 \dd{t_{\hspace{-0.4pt}\vcol{V}}}\, \dd{t_{\!\hcol{H}}}}{
      \prod\nolimits_{i = \hcol{H}, \vcol{V}} \int_{0}^{\infty} \ee^{-\Gamma t_i} \tr\!\left(\mcol{K}_i(t)^\dagger \mcol{K}_i(t)\, \mcol{\rho_M}\right) \dd{t_i}
    }.
  \end{gathered}
\end{equation}

According to \cref{eq:two-photon-symmetric-fidelity}, a difference in $K_V(t_V)$ and $K_H(t_H)$ will generically give rise to a reduction in fidelity. This can occur because of two distinct reasons: first, there might e.g.~be a difference in geometry between the decay channels, such that $K_V$ and $K_H$ are fundamentally different (discussed in \cref{sec:two-photon-geometry-errors}). But even if the decay channels are perfectly matched, an unaccounted-for difference in detection times still can be an intrinsic source of errors (\cref{sec:two-photon-timing-errors}).

\subsection{Timing-related errors}
\label{sec:two-photon-timing-errors}

The detection times $t_H$ and $t_V$ of the two herald photons are both independently distributed, as given by the exponential envelope determined by the excited state lifetime $\Gamma^{-1}$. Even if any geometric differences between the kick operators are neglected, $K \defeq K_H \approx K_V$, consider $K(t_H)$ and $K(t_V)$ associated with the two times offsets within the wavepacket (\ref{eq:two-photon-wavepacket}). In the interaction-picture phase space for the $j$-th motional mode with frequency $\motfreqelem$ (see \cref{fig:two-photon-timing}b), the first momentum kick from the excitation laser $L$ is identical for both $\ket{0}$ ($H$) and $\ket{1}$ ($V$) components, but the direction of the subsequent emission kick is entangled with the photon wavepacket. A difference $\Delta t \defeq t_V - t_H \neq 0$ in detection times leads to an angle of $\mu_j \Delta t$ between the displacement vectors.

Consequently, detecting photons at different times leaves behind entanglement between internal states and motion, dephasing the entangled state after tracing out the motional degrees of freedom. \Cref{fig:two-photon-timing}c shows numerical results for \na{}~0.6 collection from two representative configurations for various initial thermal states. To obtain an analytical approximation, consider again the zero-\na{} limit, where $w_{i\sigma}(\vvn{k}) \propto \delta(\vvn{k} - \vvn{k}_{\mathrm{coll}})$ and the total effect of $K(t)$ is a displacement. Evaluating \cref{eq:two-photon-symmetric-fidelity} for an initial thermal state with mean occupation numbers $\nbar_{th, j}$ in each mode $j$ then gives:
\begin{equation}
  \label{eq:two-photon-timing-small-na}
  C = \Gamma^2 \int_{0}^{\infty} \ee^{-\Gamma t} \int_{0}^{\,t} \prod\nolimits_j \ee^{-2 \hat{k}_{\mathrm{coll},j}^2 \lambdicke{j}^2\, (1 - \cos(\motfreqelem \Delta t)) (1 + 2 \nbar_{th, j}) }\dd{\Delta t}\, \dd{t}
\end{equation}
Assuming further that the trap potential is isotropic ($\motfreqelem \eqndef \mu$, $\lambdicke{j} \eqndef \eta$ for all $j$), and $\mu \ll \Gamma$ (such that $\motfreqelem \Delta t$ remains small in good approximation), this is approximately:
\begin{equation}
  C \approx \ee^{-2\, \eta^2 \left(\mu / \Gamma\right)^2 (1 + 2 \nbar_{th})}.
\end{equation}

Note that for parameters typical of trapped-ion experiments (e.g.~\citeref{nadlingerDeviceIndependentQuantumKey2021}), the dephasing error exceeds $\SI{0.5}{\percent}$ not too far from the Doppler-cooling limit, so is non-negligible given state-of-the-art fidelities. From \cref{eq:two-photon-timing-small-na}, it is clear that the dephasing can be reduced by rejecting events with a detection time difference $\Delta t$ beyond some cutoff, or by ground-state-cooling the atoms before excitation. Both these approaches reduce the entangled state generation rate, however. Instead, we propose to correct the resulting spin–motion entanglement through a coherent operation after the herald event takes place. Concretely, in analogy to \cref{eq:single-photon-correction} for the single-photon protocol, once the photon detection times $t_H$ and $t_V$ are known, both nodes apply
\begin{equation}
  U_{\mathrm{corr}}(t_H, t_V) = \proj{0} \otimes M_{\mathrm{corr},H}(t_H) + \proj{1} \otimes M_{\mathrm{corr},V}(t_V),
\end{equation}
where the spin-dependent corrections are chosen to maximise $C$ in \cref{eq:two-photon-symmetric-fidelity} after $K_H(t_H) \mapsto M_{\mathrm{corr},H}(t_H)\, K_H(t_H)$, $K_V(t_V) \mapsto M_{\mathrm{corr},V}(t_V)\, K_V(t_V)$, that is, aiming to disentangle the internal and motional states. In the zero-\na{} case, this can be achieved exactly by a displacement to overlap the motional states again (see \cref{fig:two-photon-timing}b). For typical experimental parameters, even such a simple spin-dependent displacement correction still significantly suppresses the error (see \cref{fig:two-photon-timing}c); for instance, for $\eta = 0.07$, $\mu / \Gamma = 0.1$ (typical of prior experiments with \srplus{} ions \cite{stephensonHighRateHighFidelityEntanglement2020,nadlingerDeviceIndependentQuantumKey2021}), the error at $\nbar_{th} = 20$ is reduced from $\num{2e-3}$ to $<\num{1e-5}$, and at $\nbar_{th} = 100$, from $\SI{1}{\percent}$ to $\SI{0.01}{\percent}$.

\subsection{Geometrical errors}
\label{sec:two-photon-geometry-errors}

\begin{figure}
  \includegraphics{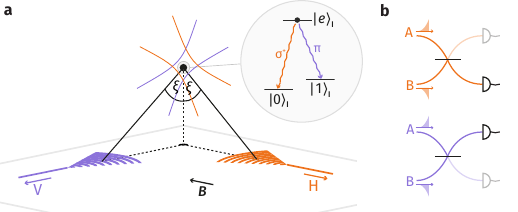}
  \includegraphics{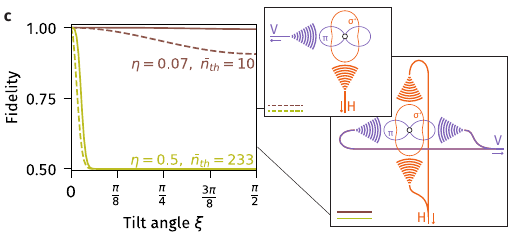}
  \caption{Two-photon-based entanglement with two modes collected from different directions.
  \figpart{(a)} An example scenario, where an atom is trapped above a planar device and polarisation-encoded photons are collected into integrated waveguide optics using grating couplers. Separating the polarisations into separate waveguides increases efficiency and lessens technical demands, but the angular offset $\xi$ of the modes from a shared axis gives rise to a differential phase sensitivity. $\vv{B}$ indicates the atomic quantisation direction.
  \figpart{(b)} The emission from two such zones is measured in the Bell basis, which can be implemented conventionally as in \cref{fig:two-photon-timing}a, or separated by polarisation.
  \figpart{(c)} Reduction in fidelity for experimental parameters characteristic of trapped-ion (brown) and optical tweezer (olive) experiments, assuming an isotropic trap with Lamb–Dicke parameter $\eta$ for all directions, as a function of the angle $\xi$ that the collection cones make with the surface normal (thermal state of motion, zero-\na{} limit). The dashed lines show the Bell state fidelities for a pair of collection gratings from a single side; the solid lines the significantly improved performance from coherently summing the outputs from two symmetrically-arranged collection devices.
  }
  \label{fig:two-photon-geometry}
\end{figure}

In the previous section, we set $K_H(t) = K_V(t)$ to focus solely on timing-related effects. In general, the position dependence/recoil effects may of course differ between the decay channels. As atom–photon entanglement is typically generated with two narrowband decay channels of similar absolute frequency and hence photon momentum, such differences stem from different geometric weights $w_{i\polind}(\vvn{k})$ in \cref{eq:all-in-one-op-def}. For the situation common to previous experimental demonstration and illustrated in \cref{fig:collection}, where a polarisation-insensitive optical system images two decay channels into a single-mode fibre,
such effects are small: even for collection of $\pi$/$\sigma$ radiation perpendicular to the magnetic field, where the coupled components correspond to two linear dipoles orthogonal to the collection axis at right angles to each other~\cite{stephensonHighRateHighFidelityEntanglement2020,nadlingerDeviceIndependentQuantumKey2021},
the error from the angular asymmetry of the dipole emission pattern away from its plane of symmetry is very small at realistic parameters (a numerical evaluation of this, as well as polarisation mixing from imperfectly suppressed longitudinal dipoles, is given in \cref{sec:high-na-orthogonal-errors}).

In this section, we instead illustrate the versatility of our framework using a different, technologically relevant case: that where emission is collected using different optical paths for the two decay channels. This is of particular interest in the context of waveguide photonics integrated into microfabricated ion-trap chips~\cite{mehtaIntegratedOpticalAddressing2016,mehtaIntegratedOpticalMultiion2020}%
, where device limitations can make it preferable to collect the two polarisation components separately~\cite{knollmannIntegratedPhotonicStructures2024,smedleyAtomicFluorescenceCollection2025}. A possible realisation is schematically depicted in \cref{fig:two-photon-geometry}a; as long as the different waveguides are stable in optical length relative to each other, the two polarisations then never need to be transmitted in the same waveguide to realise the Bell-basis measurement~\cite{ainleyMultipartiteEntanglementMultinode2024,smedleyAtomicFluorescenceCollection2025}.

For moving emitters, however, this introduces a new source of motional errors, which again admits a description in two complementary ways: in the position-dependent-phase picture, motion can now result in a differential phase shift between the components depending on its alignment with the collection directions, while in the phase space recoil picture, the (signed) magnitude of the momentum kick for each of the modes will now differ between the channels. Either way, this leads to unwanted spin–motion entanglement, and hence to errors.

\Cref{eq:two-photon-symmetric-fidelity} still captures these effects for now appropriately modified kick operators $K_H$ and $K_V$ to reflect the different collection modes. Concretely, consider the scenario from \cref{fig:two-photon-geometry}a with collection from directions each tilted an angle $\xi$ in orthogonal directions from what would be a shared collection axis, and for simplicity, assume an isotropic trap with motional frequency $\mu$ and an initial thermal state with occupation number $\nbar_{th}$. In the zero-\na{} approximation, and ignoring the time-dependent effects already described in \cref{sec:two-photon-timing-errors} (i.e.~taking $\Gamma \rightarrow \infty$), we find
\begin{equation}
  \label{eq:two-photon-geometry-contrast}
  C = \ee^{-2 \gamma}\ ,\quad \textrm{with} \quad \gamma \defeq \eta^2 (\sin \xi)^2 (1 + 2 \nbar_{th}).
\end{equation}
This is illustrated for typical experimental parameters in \cref{fig:two-photon-geometry}c (dashed curves). Once again, an (approximate) correction of these effects through appropriate spin-dependent operations chosen to to disentangle internal and motional states would be viable.

Alternatively, however, as recently also proposed in \citeref{smedleyAtomicFluorescenceCollection2025}, a passive improvement is possible as well: by coherently combining the emission from two couplers symmetrically placed around the atom location (see \cref{fig:two-photon-geometry}c), the emission is effectively collected into a standing wave pattern in the trap plane (though it is still a running wave in the perpendicular direction common to both collection directions). Even in the zero-\na{} limit, $w(\vvn{k}) \propto (\delta(\vvn{k} - \vvn{k}_{\mathrm{coll}}) + \delta(\vvn{k} + \vvn{k}_{\mathrm{coll}})) / \sqrt{2}$, the collection efficiency now drops as $(1 + \ee^{-2 \gamma}) / 2$, and the contrast from \cref{eq:two-photon-geometry-contrast} instead becomes
\begin{equation}
  C = \frac{4 \ee^{-2\gamma}}{(1 + \ee^{-2\gamma})^2} = (\operatorname{sech} \gamma)^2.
\end{equation}
For small $\gamma$, the Bell state error $1 - F$ in the one-sided case is $\gamma + O(\gamma^2)$, while the two-sided collection reduces it to $\gamma^2 / 2 + O(\gamma^4)$.

This reduction in error emphasises that the correct interpretation of the weights $w_{i\polind}(\vvn{k})$ for free-space collection into a single mode is not one of an incoherent mixture of different recoil directions, but in fact a coherent superposition. Phase-stable operation of standing waves generated by grating couplers integrated into ion traps was recently demonstrated~\cite{vasquezControlAtomicQuadrupole2023,hattoriIntegratedPhotonicsBasedArchitecturesPolarizationGradient2022}, so this could be a valuable technique to reduce errors in intra- or inter-chip photonic links using integrated optics.

\section{Two-photon-based entanglement using~time-bin encoding}
\label{sec:time-bin-encoding}

\begin{figure}
  \begin{minipage}[ht!]{0.48\linewidth}
    \includegraphics{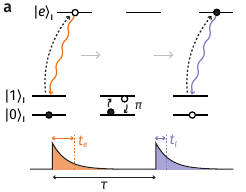}
    \includegraphics{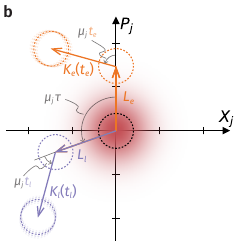}
  \end{minipage}\hfill%
  \begin{minipage}[ht!]{0.52\linewidth}
    \includegraphics[width=\linewidth]{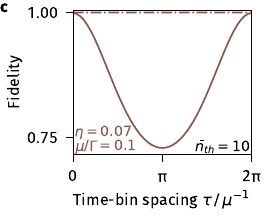}
    \vspace{6pt}
    \includegraphics[width=\linewidth]{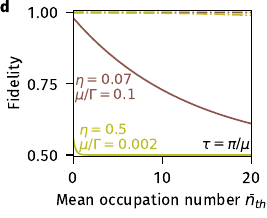}
  \end{minipage}%
  \vspace{-6pt}
  \caption{Time-bin-encoded two-photon-heralded entanglement.
    \figpart{(a)} To generate entanglement between internal states and photonic time bins, the atom is first prepared in an equal superposition of $\ket{0}_I$ and $\ket{1}_I$, and only the $\ket{1}_I$ is excited to the emitting state $\ket{e}_I$. The states are then swapped using a $\pi$ pulse, and the process is repeated. A partial Bell basis measurement (not pictured) on a pair of such maximally entangled states $(\ket{0e}_{IF} + \ket{1l}_{IF}) / \sqrt{2}$ in the heralding station yields a maximally-entangled atom–atom state.
    \figpart{(b)} As illustrated in the phase space of the $j$-th motional mode of one of the nodes, there is a differential displacement between the $e$arly and $l$ate time bins, depending now not only on $t_e$ and $t_l$, but also on the time-bin spacing $\tau$.
    \figpart{(c)} The amount of spin-motion entanglement in mode $j$ depends on $\motfreqelem$; it is maximal when $\tau$ is such that the excitation kicks $L_e$ and $L_l$ happen in opposite directions in the rotating frame, leading to a minimum in fidelity (solid line). A spin-displacement correction (dash-dotted line) recovers virtually all the loss in fidelity (shown for collection into a Gaussian mode with numerical aperture 0.6, excitation perpendicular to the collection direction, an isotropic trap, and a thermal state).
    \figpart{(d)} For the worst time-bin spacings, where $\mu \tau$ is an odd multiple of $\pi$, the loss in fidelity is considerable even for near-ground-state-cooled atoms (solid lines), but is efficiently suppressed by a spin-dependent post-herald displacement (dash-dotted lines).
  }
  \label{fig:time-bin-encoding}
\end{figure}

By initialising the atoms into an equal superposition of states and subsequently exciting each of the components to the same short-lived emitting state decaying on a closed transition (separated by a $\pi$ pulse to swap the components, see \cref{fig:time-bin-encoding}a), the two-photon protocol can be realised using time bins rather than polarisation or frequency for the photonic encoding~\cite{barrettEfficientHighfidelityQuantum2005}.
This can be desirable because of the relative insensitivity of such protocols to common link disturbances such as birefringence, or for extension to higher-dimensional quantum systems (qudits with $d > 2$).
In contrast to solid-state emitters, e.g.~nitrogen vacancy centres, where the robustness of this encoding scheme allowed fidelities $>\SI{90}{\percent}$~\cite{bernienHeraldedEntanglementSolidstate2013,Hensen2015} to be reached, motion complicates the picture for trapped atoms.

Broadly, the analysis for two-photon protocols from \cref{sec:two-photon-errors} is also applicable here, with the difference that the kick operators corresponding to the qubit state $\ket{0}_I$ and $\ket{1}_I$, which we shall label $K_e(t)$ (early) and $K_l(t)$ (late), now correspond to excitation laser kicks and subsequent emission separated by the time-bin spacing $\tau$. In the rotating-frame phase space for each mode $j$, the displacements from excitation laser kick and decay for the late time-bin are now rotated by $\mu_j \tau$ (see \cref{fig:time-bin-encoding}b). Thus, even if $\Gamma \gg \mu$, such that the spread in decay times is negligible, the momentum kicks from the excitation laser alone can already leave behind spin-motion entanglement. This is similar to the single-photon protocol discussed in \cref{sec:single-photon}, though here effect is periodic in $\tau$, as the late time-bin direction rotates in and out of alignment with that for the early time bin.

Concretely, take the two photon detection times $t_e$ and $t_l$ to be relative to the start of their time bin, and consider an excitation laser pulse along $\vvn{k}_{ex}$. The kick operators are then
\begin{equation}
  \begin{aligned}
    K_{e}(t_e) &\defeq \int_{S^2} w(\vvn{k})\, \dmulti{-\ii \vvn{k}}{t_e}\,\ddtwo{\vvn{k}}\ \cdot \ \dmulti{\ii \vvn{k}_{ex}}{0},\\
    K_{l}(t_l) &\defeq \int_{S^2} w(\vvn{k})\, \dmulti{-\ii \vvn{k}}{\tau + t_l}\,\ddtwo{\vvn{k}}\ \cdot \ \dmulti{\ii \vvn{k}_{ex}}{\tau},\\
  \end{aligned}
\end{equation}
and evaluating \cref{eq:two-photon-symmetric-fidelity} in the zero-\na{} limit, $w(\vvn{k}) \propto \delta(\vvn{k} - \vvn{k}_{\mathrm{coll}})$, for nodes with identical properties, we obtain the fidelity as $F = (1 + C) / 2$,
\begin{equation}
  C = \Gamma^2 \int_{0}^{\infty}\int_{0}^{\infty} \ee^{-\Gamma (t_e + t_l)} \prod\nolimits_j \ee^{-2 \lambdicke{j}^2\, \absval{\beta_j(t_e, t_l)}^2\, (1 + 2 \nbar_{th, j}) }\dd{t_l}\, \dd{t_e},
  \label{eq:time-bin-fidelity-zero-na}
\end{equation}
where
\begin{equation}
  \beta_j(t_e, t_l) \defeq \kexelem (1 - \ee^{\ii \motfreqelem \tau}) - \hat{k}_{\mathrm{coll},j} (\ee^{\ii \motfreqelem t_e} - \ee^{\ii \motfreqelem (t_l + \tau)})
  \label{eq:time-bin-residual-displacement}
\end{equation}
quantifies the amount of residual phase space displacement in mode $j$. If we, for clarity of exposition, assume the confinement to be symmetric and $\kex$ to be perpendicular to $\vvn{k}_{\mathrm{coll}}$, this further simplifies to
\begin{equation}
  C = \frac{\Gamma^2}{2} \int_{0}^{\infty} \ee^{-\Gamma t} \int_{-t}^{\,t} \ee^{-2 \eta^2\, (2 - \cos(\mu \tau) - \cos(\mu (\Delta t + \tau))) (1 + 2 \nbar_{th}) }\dd{\Delta t}\, \dd{t},
\end{equation}
which is similar to the non-time-bin two-photon case (\ref{eq:two-photon-timing-small-na}), but $\tau$ offsets the arrival time difference $\Delta t \defeq t_l - t_e$, and the excitation pulses reduce the contrast by an extra $\ee^{-2 \eta^2 (1-\cos(\mu \tau))(1 + 2 \nbar_{th})}$ even if $\Gamma \rightarrow \infty$ (solid lines in \cref{fig:time-bin-encoding}c–d). This can be avoided by choosing the time-bin spacing to be an integer multiple $n$ of the trap period, $\mu \tau = 2\pi\,n$, but to simultaneously meet this criterion for all motional modes at both nodes at the same time may be a cumbersome constraint.

Instead, as discussed in the previous sections, the recoil effect can also be cancelled by the application of a spin-dependent force after heralding: if the atom is excited to $\ket{e}_I$ using an approximately plane-wave-like laser beam, the excitation terms in the residual displacement \cref{eq:time-bin-residual-displacement} are known exactly. Depending on $\mu / \Gamma$, the displacement to correct also depends on the relative timing of $t_e$ and $t_l$ in the same way as discussed in \cref{sec:two-photon-timing-errors}. Outside the zero-\na{} limit, this arrival time can only approximately be corrected by separable spin-dependent displacements; nevertheless, this brings the time-bin encoded schemes to the same level of error sensitivity as e.g.~polarisation-encoding schemes. As shown in \cref{fig:time-bin-encoding}d, the spin-dependent displacement correction restores the fidelity at $\eta = 0.07$, $\mu / \Gamma = 0.1$, $\nbar_{th} = 20$ (fibre-coupled using \na{} 0.6) from $\SI{61}{\percent}$ to $>\SI{99.99}{\percent}$ even at the most unfavourable time-bin duration. (This is a feed-forward correction of a known effect, in contrast to a recent proposal for rephasing of unknown energy-level shifts~\cite{uysalRephasingSpectralDiffusion2024}.)

In situations where the robustness of time-bin protocols to polarisation effects, or their extensibility to higher-dimensional systems, is desirable, our results thus provide a practical guideline towards high-fidelity operation.

\section{Conclusion}

Our analysis yields collection efficiencies and entangled-state fidelities for a wide variety of protocols and collection geometries: it can be specialised to the case of a vanishing collection angle and thermal states to obtain analytical approximations, but seamlessly generalises to any arbitrarily-structured collection modes, such as imaging into single-mode fibres or multiple combined waveguide couplers, in the high-\na{} limit relevant to technological applications aiming for maximal collection efficiency.
We have focussed on bipartite entanglement generation between qubits, as this is the most well-explored use case in theory and experiment, but the core kick-operator framework describing the atom–photon wavepackets applies also to entanglement generation between qudits and direct multipartite entanglement generation~\cite{ainleyMultipartiteEntanglementMultinode2024}.

For the practical performance of the correction techniques proposed here, motional decoherence might become relevant, as could non-ideal single-photon detectors (with finite time resolution or dark counts), all of which are readily incorporated into this model.
In future work, the model could also be augmented to describe system-specific motional dynamics, such as the micromotion of ions in radiofrequency traps (by considering a time-dependent position operator, cf.~\citeref{Leibfried2003}).
Corrections more accurate than the spin-dependent displacements discussed here appear possible (e.g.~using higher-order motional operations, such as spin-dependent squeezing, or multi-mode operations); we leave the question of which kick operators can be corrected exactly to future work.

More speculatively, we note that while the current discussion focussed on the deleterious consequences of the atomic motion for heralded entanglement generation, fundamentally the model describes a heralded application of a spin–motion operator. Thus, the same mechanism might be leveraged to generate complex spin–motion interactions heralded by the detection of a single photon, e.g.~in an interestingly-structured light mode, or to mediate interactions between the internal/motional states of multiple interacting emitters.

In summary, we have introduced a flexible framework to describe the role of motion in spontaneous-emission-mediated entanglement generation between trapped atomic qubits. A set of \enquote{kick operators} succinctly encapsulates the geometric aspects of the problem, linking together internal, motional and photonic degrees of freedom. While recoil leads to errors universal to all trapped optical emitters, they can be suppressed by reducing the residual spin-motion entanglement after a herald event. We have illustrated simple correction strategies for several entanglement generation protocols; as they only have to be—indeed, can only be—applied after a successful herald, the impact on the entanglement generation rate is minimal. Thus, our results yield improvements in high-rate, high-fidelity remote entanglement generation, whether through relaxation of cooling requirements for high-fidelity operation, through increased flexibility in the use of time-bin encoding, or through enabling the use of single-photon schemes for large rate improvements in link-loss-dominated long-distance settings.

\vspace{20pt}

\emph{Note:} In complementary work, Kikura et al.~(pre-print: ref.~\cite{kikuraTamingRecoilEffect2025}) apply the kick operator framework to cavity-assisted photon collection and detail the implications for time-bin photonic entanglement generation between neutral atoms in optical tweezer arrays. During preparation of this manuscript, we also became aware of a recent experimental demonstration of high-fidelity time-bin-mediated entanglement between trapped ions by Saha et al.~\cite{sahaHighfidelityRemoteEntanglement2025}, who give error estimates similar to the time-averaged zero-\na{} case presented in \cref{sec:time-bin-encoding}.

\begin{acknowledgments}
  The authors would like to thank the members of the Oxford Ion Trap Quantum Computing group (in particular R.\,Srinivas, G.\,Araneda, P.\,Drmota, and D.\,Main) and J.\,Home for inspiring discussions. We are grateful to A.\,Martinez, P.\,Drmota, and D.\,M.\,Lucas for helpful comments on the manuscript.
  D.\,P.\,N.~acknowledges Merton College, Oxford for support through a Junior Research Fellowship. 
\end{acknowledgments}

\clearpage
\onecolumngrid
\appendix

\section{Spontaneous decay derivation}
\label{sec:decay-derivation}

\newcommand{\ketketket}[3]{\ket{#1}\hspace{-1pt}\ket{#2}\hspace{-1pt}\ket{#3}}
\newcommand{\nvec}{{\vec{n}}}
\newcommand{\lvec}{{\vec{l}}}

\begin{figure*}
	\includegraphics{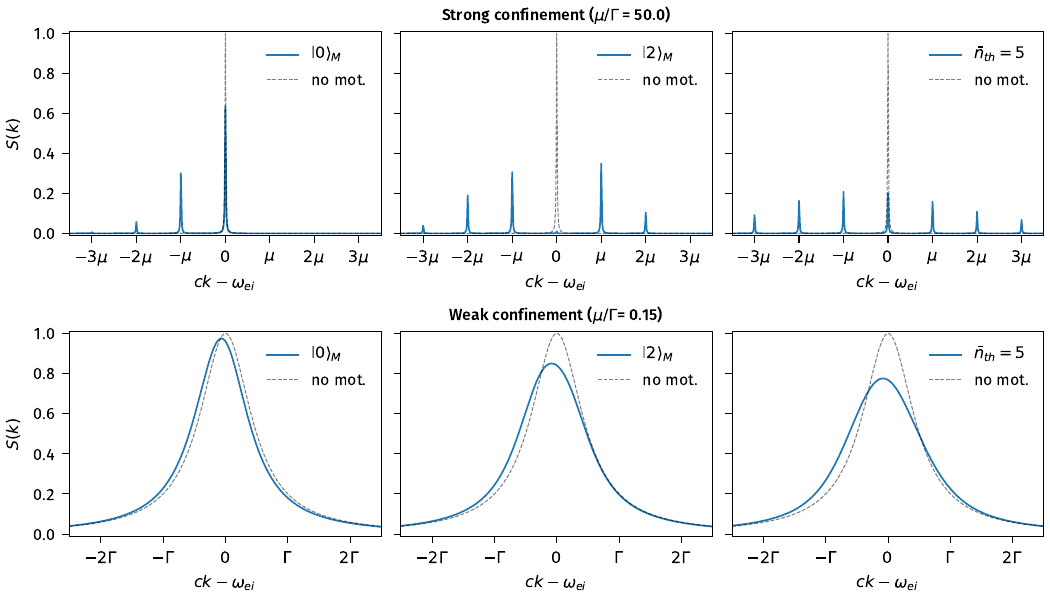}
	\caption{Spectral energy density of spontaneous emission from a trapped emitter in one dimension. A large Lamb–Dicke parameter of $\eta=1.0$ exaggerates the effect for illustration purposes.}
	\label{fig:decay-spectrum}
\end{figure*}

In this section, we derive the pure state describing the internal state of the atom, its motion, and the emitted photon after spontaneous decay from a harmonic trap. We start with the interaction picture Hamiltonian \eqref{eq:hint-rwa}, looking to solve for the evolution of the initial state $\ket{\psi_0}_{IMP} = \ket{e}_I\ket{\zeta}_M\ket{\vac}_P$. For a single atom, we can expand its motional state in the 3D Fock basis, $\ket{\nvec} = \ketketket{n_1}{n_2}{n_3}$, but this expansion can be generalised to a different number of modes (e.g. multiple ions in the same trap).
In line with the standard spontaneous emission derivation, we then assume an (interaction picture) Ansatz with an excited, motionally unchanged contribution and a decayed, disturbed, contribution with a photon:
\begin{equation}
	\ket{\psi(t)} = \sum_\nvec \alpha_\nvec(t) \ket{e}_I \ket{\nvec}_M \ket{\vac}_P
	+ \sum_\nvec \sum_{i \modeind} \beta_{i \modeind \nvec}(t) \ket{i}_I \ket{\nvec}_M \adag{\modeind}\ket{\vac}_P.
\end{equation}
Plugging this Ansatz into the interaction picture Schr{\"o}dinger equation, we obtain coupled first order equations for the excited and decayed amplitudes
\begin{align}
	\hbar \frac{\dd \alpha_\nvec}{\dd t}
	&=
	\sum_{i \modeind}
	\ee^{\ii(\omega_{ei}-kc)t} \fieldcpl \overline{\statecpl{i}} \sum_\lvec \beta_{i \modeind \lvec}(t)
	\braopket{\nvec}{\dmulti{\ii\vvn{k}}{t}}{\lvec}
	\\
	\hbar \frac{\dd \beta_{i \modeind \nvec}}{\dd t}
	&=
	- \ee^{\ii(kc-\omega_{ei})t} \fieldcpl \statecpl{i} \sum_\lvec \alpha_\lvec(t)
  \braopket{\nvec}{\dmulti{-\ii\vvn{k}}{t}}{\lvec}\, .
  \label{eq:decayedcoeff}
\end{align}
Next, as in the Wigner-Weisskopf treatment \cite{weisskopfBerechnungNaturlichenLinienbreite1927,Scully1997}, we transition to continuum modes, make the Markovian approximation for $\alpha_\nvec(t)$, assume the radiation is closely distributed about the transition frequency $\omega_{ei}$, and extend the integration bounds over negative frequencies $k$. Then, assuming the Lamb shift is already included in the energy spacings, we recover the exponential decay law for the excited coefficients, $\alpha_\nvec(t) = \ee^{-\frac{\Gamma}{2}t} \braket{\nvec}{\zeta}$. With this solution in hand, we can solve for the amplitudes related to emission through integration of \eqref{eq:decayedcoeff}:
\begin{equation}
	\beta_{i \modeind \nvec}(t)
	=
	- \frac{\fieldcpl \statecpl{i}}{\hbar}  \int_0^{\, t}
	\ee^{\ii t'(kc - \omega_{ei} + \ii \Gamma/2)}
  \braopket{\nvec}{\dmulti{-\ii\vvn{k}}{t'}}{\zeta} \, \dd{t'} \, .
\end{equation}
Finally, we insert a resolution of the identity and take the time evolution of the displacement operator outside of the matrix element to find the state after decay.
\begin{equation}
	\label{eq:freqdomainstate}
	\ket{\psi(t)} =
	\;  \ee^{-\Gamma t /2}
	\ketketket{e}{\zeta}{0}
	 -
	\sum_{i\sigma\nvec\lvec}\,
	\int_{\mathbb{R}^3}
	\frac{\fieldcpl \statecpl{i}}{\hbar} \braketf{\lvec}{\zeta}
	\frac{
    \braopketf{\nvec}{\dmulti{-\ii\vvn{k}}{0}}{\lvec}
		 \left(1 - \ee^{-\Gamma t/2} \ee^{\ii t \left(
			kc - \omega_{ei} - \vec{\mu}\cdot(\lvec - \nvec)
			\right)
			}\right)
	}{
		\Gamma/2 + \ii \left(
		\omega_{ei}  + \vec{\mu} \cdot (\lvec - \nvec)
		-ck
		\right)
	}
	\ket{i}\ketf{\nvec} \adag{\modeind} \ket{\vac}
  \ddthree{\vv{k}}
\end{equation}
After waiting much longer than the decay lifetime, $t \gg \Gamma^{-1}$, we are left with an interaction-picture state approximately constant in time. From the denominator of the photon amplitude, we can see that the emission spectrum will consist of superposed Lorentzians with width $\Gamma$ centred at $ck = \omega_{ei} + \vec{\mu}\cdot(\lvec - \nvec)$. The relative weights of these peaks depend on the initial motional state. To illustrate the resulting spectrum, \cref{fig:decay-spectrum} shows the spectral energy density for both strong and weak confinement in one dimension.
In the strong confinement (or resolved-sideband) regime $\mu  /  \Gamma \gg 1$, transitions between motional states are clearly defined, and conservation of energy leads to peaks in emitted photon wavenumbers spaced by $\mu$. On the other hand, in the weak confinement regime practical for remote entanglement generation between atoms, the motional changes are less distinguishable. The Lorentzians overlap significantly, and the effect on the spectrum amounts to broadening.
\par{}
To better interpret the properties of the post-emission state \eqref{eq:freqdomainstate} and to obtain further simplifications, we can rewrite it in terms of time-domain modes
, $\adag{\sigma}(t,\vvn{k}) = \sqrt{\frac{c}{2\pi}} \int \adag{\sigma}(k,\vvn{k}) \ee^{\ii k c t}\, \dd k $. Making repeated use of the narrowband approximation, $\Gamma \ll \omega_{ei}$, the state in the long-time limit may be written as
\begin{equation}
	\ket{\psi}_{IMP} \approx 
	\sum_i
	\ket{i}_I
  \fieldcplcontcentre{i}
	\int_0^\infty
	\ee^{-({\Gamma}/{2} + \ii \omega_{ei})t} \\
	\int_{\mathcal{S}^2}
	\sum_\sigma
	\statecpl{i} \,
  \dmulti{-\ii\vvn{k}}{t}
	\ket{\zeta}_M
	\adag{\sigma} (t,\vvn{k}) \ket{\vac}_P
	\ddtwo{\vvn{k}} \,
	\dd{t}
\end{equation}
with an adjusted prefactor $\fieldcplcontcentre{i} \defeq - \sqrt{\frac{2\pi}{\hbar^2 c}}\mathcal{E}_{\omega_{ei} / c} \frac{\omega_{ei}^2}{c^2}$, and the second integral is over the unit sphere. Note that we are still in the interaction picture where this state is time-independent; $t$ now denotes the time coordinate within the photonic wavepacket.

\section{Coupling into a Gaussian fibre mode}
\label{sec:gaussian-mode-calculations}
To describe the coupling of emitted radiation into the output mode from each node, for instance the approximately Gaussian mode of a single-mode fibre, we are interested in calculating the overlap coefficients $\alpha_{\polind}(\vvn{k})$ from \cref{sec:collection}/\cref{fig:collection}, quantifying how much each plane wave mode contributes to the fibre mode after the collection optics. The desired coefficients form part of the unitary imaging system transformation $\adag{\polind}(\vv{k}) = \alpha_{\polind}(\vvn{k})\, f_{\polind}^\dagger(k) + \ldots$, where the dots elide the modes traced over, corresponding to emission not coupled into the fibre. To achieve this in the high-\na{}, vector-optics setting, we propagate both modes to a reference sphere at the input of the imaging system~\cite{richardsElectromagneticDiffractionOptical1959,novotnyPrinciplesNanoOptics2012} (around the trap location), where we perform the mode overlap integral.

The profile of the fibre mode propagating along $\vvn{z}$ with polarisation $\sigma_F$ is given by
\begin{equation}
	\vv{\varepsilon}_{F,\sigma_F}(\rho,\phi) = \sqrt{\frac{2}{\pi w_0^2}} \ee^{-\left(\frac{\rho}{w_0}\right)^2} \left( \cos(\sigma_F)\vvn{x} + \sin(\sigma_F)\vvn{y}\right) \, ,
\end{equation} where $\rho$ and $\phi$ denote the polar coordinates of the beam. This profile is normalised over the $x$-$y$ plane and is defined up to a phase factor determined by position along the beam.

Supposing the plane wave vector of the mode $\vv{k}\polind_P$ is given in spherical coordinates, $\vv{k} = (k,\vartheta,\varphi)$, the electric field profile at position $\vv{r} = (r,\theta,\phi)$ is given by
\begin{equation}
	\vv{\varepsilon}_{P,\vv{k}\polind_P}(\vv{r}) = \ee^{\ii \vv{k} \cdot \vv{r}} \vvn{\epsilon}_{\vvn{k}\polind_P}(\vartheta,\varphi)
\end{equation}
for a given polarisation vector $\vvn{\epsilon}_{\vvn{k}\polind_P}$.

Consider the lens of focal length $f$ to be positioned along the $z$ axis, with focus at the coordinate origin. It then transforms the incoming Gaussian mode as $\vvn{r} \leftrightarrow \vvn{\theta}$, $\rho = f \sin\theta$, and an apodisation factor $\sqrt{\cos\theta}$ \cite[ch.~3]{novotnyPrinciplesNanoOptics2012}. Both field profiles on the reference sphere $r=f$ are then given by
\begin{align}
	\vv{\varepsilon}'_{F,\sigma_F}(f\vvn{r}) = \sqrt{\frac{2}{\pi w_0^2}} \sqrt{\cos \theta} \, \ee^{-\left(\frac{f \sin \theta}{w_0}\right)^2}
	\left( \cos(\sigma_F-\phi)\vvn{\theta} + \sin(\sigma_F-\phi)\vvn{\phi}\right)
	& & \mathrm{and} & &
	\vv{\varepsilon}_{P,\vv{k}\polind_P}(f\vvn{r}) = \ee^{\ii kf \vvn{k} \cdot \vvn{r}} \vvn{\epsilon}_{\vvn{k}\polind_P}(\vartheta,\varphi) \, .
\end{align}
Now we are ready to calculate the overlap integral over the reference sphere, where the polar angle goes up to $\vartheta_{max}$ (\na{} $\sin{\vartheta_{max}}$). The desired coefficient is given by
\begin{align}
	\alpha(\vv{k}) = \int_0^{\vartheta_{max}} \int_0^{2\pi}
	\left(
	\overline{\vv{\varepsilon}'_{F,\sigma_F}(f\vvn{r})}
	\cdot
	\vv{\varepsilon}_{P,\vv{k}\polind_P}(f\vvn{r})
	\right)
	 \sin(\theta) \, \dd \phi
	 \, \dd \theta \, .
\end{align}
Performing the integral over $\phi$, we arrive at
\begin{equation}
	\alpha(\vv{k}) =
	\sqrt{\frac{8\pi}{w^2_0}}
	\int_0^{\vartheta_{max}} \sin(\theta) \sqrt{\cos(\theta)} \, \ee^{-\left(\frac{f \sin \theta}{w_0}\right)^2} \ee^{\ii f k \cos \vartheta \cos\theta}
	\begin{pmatrix}
		\cos^2(\theta/2) \cos \sigma_F J_0(\xi) + \sin^2(\theta/2) \cos(2\varphi - \sigma_F) J_2(\xi)
		\\
		\cos^2(\theta/2) \sin \sigma_F J_0(\xi) + \sin^2(\theta/2) \sin(2\varphi - \sigma_F) J_2(\xi)
		\\
		- \ii \sin \theta \cos(\varphi - \sigma_F) J_1(\xi)
	\end{pmatrix}
	\cdot \vvn{\epsilon}_{\vvn{k}\polind_P}(\vartheta,\varphi)
	\, \dd \theta
	\label{eq:couplingcoeff}
\end{equation}
having defined $\xi = fk \sin \vartheta \sin \theta$ for convenience.
\par{}
This expression for the mode coupling coefficients contains highly oscillatory integrals from diffraction on the aperture stop,
\begin{equation}
	\mathcal{I}_\nu(\vartheta) \defeq \int_0^{\vartheta_{max}} \ee^{\ii f k \cos \vartheta \cos \theta} J_\nu(\xi) \, S(\theta) \, \dd \theta
\end{equation}
for $fk \gg 1$, $\nu = 0, 1, 2$, and slowly varying, well-behaved functions $S(\theta)$. By the method of stationary phase \cite[ch.~6]{benderorszag1978}\cite{ErdelyiA.1955ARoF}, such an integral can be written in the following asymptotic form for $fk \rightarrow \infty$:
\begin{align}
	\mathcal{I}_\nu(\vartheta) \approx
	\begin{cases}\displaystyle
		\frac{1}{fk} \frac{ S(\vartheta)}{\sin \vartheta} \ee^{\ii \left( fk + \frac{\pi}{2}(3\nu - 1)\right)}
		+ \mathcal{O}\left((fk)^{-3/2}\right) & 0 < \vartheta < \vartheta_{max}
		\vspace{5pt}
		\\ \mathcal{O}\left((fk)^{-3/2}\right) & \vartheta_{max} < \vartheta < \frac{\pi}{2} \, .
	\end{cases}
	\label{eq:integral}
\end{align}
If necessary, the $\mathcal{O}\left((fk)^{-3/2}\right)$ terms can be evaluated through integration by parts; they produce an oscillation on top of the slowly varying term from the stationary phase approximation, but these diffraction effects can usually be ignored. By plugging \eqref{eq:integral} into \eqref{eq:couplingcoeff}, we can therefore obtain the coupling coefficients linking atom emission in the plane-wave basis to the output in a single-mode fibre.

\section{Single-photon scheme general post-herald state}
\label{sec:single-photon-post-herald-derivation}

\newcommand{\efficiency}{E}

In the single-photon scheme, the atoms at both nodes are weakly excited, and (part of) each emitted photon wavepacket collected and interfered on a beamsplitter. A detector click then heralds the creation of an entangled state of the two atoms. In this section, we give the expressions for the post-herald state of the atoms and motion and the resulting fidelity of the scheme.

The internal structure of the two atoms is assumed to be identical and the only relevant decay route considered is $e\rightarrow 1$. Additionally, we only consider the radiation coupled to one polarisation in the single mode fibre, and pure motional states (soon to be generalised).
In the above scenario, the incoming density matrices from \cref{eq:single-photon-wavepacket} for each of the nodes $s \in \{A, B\}$ are of the form
\begin{equation}
	 \operatorname{proj}\Big( \!\sqrt{1 - p_s} \icol{\ket{0}_{I^s}} \mcol{\ket{\zeta}_{M^s}} \pcol{\ket{\vac}} + \sqrt{p_s} \icol{\ket{1}} \int_0^\infty \ee^{-\left(\frac{\Gamma}{2} + \ii \icol{\omega_{e1}}\right)\, t}
	\mcol{K_s}(t) \mcol{\ket{\zeta}_{M^s}} \pcol{\tilde{f}_s^\dagger}(t)\, \dd{t} \pcol{\ket{\vac}} \!\Big) 
	+ 
	p_s (1-\efficiency_{s}) \icol{\proj{1}_{I^s}} \otimes \mcol{ \rho_{M^s, \mathrm{dark} }} \otimes \pcol{\proj{\vac}} \, ,
\end{equation}
where the collection efficiency, $\efficiency_{s} \defeq \int_{0}^{\infty} \ee^{-\Gamma t} \tr(\mcol{K_s}(t)^\dagger \mcol{K_s}(t) \mcol{\ketbra{\zeta}{\zeta}_{M^s}} )\, \dd{t}$,
was inserted such that the post-decay, uncollected motional contribution $\rho_{M^s, \mathrm{dark} }$ is normalised. When we generalise to arbitrary motional states, we will replace $\ketbra{\zeta}{\zeta}_{M^s}$ with the relevant density operator in the definition. After combining the two nodes, noting that the photonic operators commute, we apply the beamsplitter as in section \ref{sec:single-photon}. Assuming that at time $t$ after the excitations, there was a click at the detector sensitive to mode $g$, the subnormalised state of both atoms and their motion is given by
\begin{multline}
	\rho_{\land g}(t) = \frac{\ee^{-\Gamma t}}{2} \operatorname{proj}\Bigl(
	\sqrt{1-p_A} \ket{0}_{I^A} \ket{\zeta}_{M^A} \otimes \sqrt{p_B} \ket{1}_{I^B} K_B(t) \ket{\zeta}_{M^B} 
	+ 
	\sqrt{p_A} \ket{1}_{I^A} K_A(t) \ket{\zeta}_{M^A} \otimes \sqrt{1-p_B} \ket{0} \ket{\zeta}_{M^B}
	\Bigr) \\
	+
	p_A p_B \frac{\ee^{-\Gamma t}}{2} \ketbra{1}{1}_{I^A} \ketbra{1}{1}_{I^B} \left(
	(1-\efficiency_{A})	\rho_{M^A, \mathrm{dark}} \otimes K_B(t) \ketbra{\zeta}{\zeta}_{M^B} K_B(t)^\dagger + 
	K_A(t) \ketbra{\zeta}{\zeta}_{M^A} K_A(t)^\dagger
	\otimes (1-\efficiency_{B}) \rho_{M^B, \mathrm{dark}}\right) \\
	+
	p_A p_B \frac{\ee^{-2\Gamma t}}{4} \ketbra{1}{1}_{I^A} \ketbra{1}{1}_{I^B} \left( K_A(t) \ketbra{\zeta}{\zeta}_{M^A} K_A(t)^\dagger \otimes K_B(t) \ketbra{\zeta}{\zeta}_{M^B} K_B(t)^\dagger 
	\right) \, . \label{eq:single-photon-conditional-proj}
\end{multline}
If instead detector $\tilde{h}$ went off, the only change would be the plus sign in the superposition changing to a minus. This equation illustrates sources of error in the single-photon scheme: the first line is related to the desired, maximally entangled atom–atom state, the second line contains terms from double excitation in which only one photon was collected, and the final line corresponds to double excitation and detection of two photons in mode $g$.
\Cref{eq:single-photon-conditional-proj} is straightforwardly generalised to arbitrary initial motional states $\rho_{M^A}$ and $\rho_{M^B}$ by considering their spectral decompositions $\rho_{M^A} = \sum_{m^A}p_{m^A} \proj{\zeta_{m^A}}$, $\rho_{M^B} = \sum_{m^B}p_{m^B} \proj{\zeta_{m^B}}$. Tracing out the motion and writing out the atom–atom density matrix in the computational basis, we obtain explicitly
\begin{multline}
	\rho_{\land g}(t) = \frac{\ee^{-\Gamma t}}{2} \cdot\\
	\begin{pmatrix}
		0 & 0 & 0 & 0 \\
		0 & (1-p_A)p_B \tr(K_{B}(t)\,\rho_{M^B} K_{B}(t)^\dagger) & \sqrt{p_A p_B (1-p_A)(1-p_B)} \tr(\rho_{M^A} K_{A}(t)^\dagger) \tr(K_{B}(t)\,\rho_{M^B}) & 0 \\
		0 & \sqrt{p_A p_B (1-p_A)(1-p_B)} \tr(K_{A}(t)\,\rho_{M^A}) \tr(\rho_{M^B} K_{B}(t)^\dagger)  &
		p_A(1-p_B) \tr(K_{A}(t)\,\rho_{M^A} K_{A}(t)^\dagger)  & 0 \\
		0 & 0 & 0 & p_A p_B \delta \\
	\end{pmatrix}
\end{multline}
with $\delta \defeq (1-\efficiency_{A}) \tr(K_{B}(t) \rho_{M^B} K_{B}(t)^\dagger) + (1-\efficiency_{B}) \tr(K_{A}(t) \rho_{M^A} K_{A}(t)^\dagger) + \frac{\ee^{-\Gamma t}}{2} \tr(K_{A}(t) \rho_{M^A} K_{A}(t)^\dagger) \tr(K_{B}(t) \rho_{M^B} K_{B}(t)^\dagger) $.
Allowing for all possible detection times, we can average the density matrix by integrating the post-detection state over $t$, obtaining $\rho_{\land g, \mathrm{all}} = \int_0^\infty \rho_{\land g}(t) \, \dd t$. We can then calculate the normalised atom–atom state $\rho_{| g,\mathrm{all}} =\rho_{\land g,\mathrm{all}} / \tr(\rho_{\land g,\mathrm{all}})$ and use it to get an explicit expression for the fidelity by considering the overlap with the closest Bell-like state. We obtain
\begin{equation}
	F(\rho_{| g,\mathrm{all}}) = \frac{1}{2} \frac{
	p_A \efficiency_{A} (1-p_B) + (1-p_A) p_B \efficiency_{B} + 2 \sqrt{p_A p_B (1-p_A) (1-p_B)} \absval{\int_0^\infty \ee^{-\Gamma t} \tr( K_A(t) \rho_{M^A} ) \tr(\rho_{M^B}  K_B(t)^\dagger)\, \dd{t}}}{
	p_A \efficiency_{A} + p_B \efficiency_{B} - p_A \efficiency_{A} p_B \efficiency_{B} + \frac{p_A p_B}{2} \int_0^\infty \ee^{-2\Gamma t} \tr(K_{A}(t)\,\rho_{M^A} K_{A}(t)^\dagger) \tr(K_{B}(t)\,\rho_{M^B} K_{B}(t)^\dagger)\, \dd{t}
	} \, .
\end{equation}
All of the above expressions simplify considerably if we make the assumption that both systems are identical. In that case, the fidelity simplifies to
\begin{equation}
	F(\rho_{| g,\mathrm{all}}) = \frac{1}{2}\, (1-p)\, \left. \left( 1 + \frac{\int_0^\infty \ee^{-\Gamma t} | \tr(K(t) \rho_M)|^2 \dd t}{\int_0^\infty \ee^{-\Gamma t} \tr(K(t) \rho_M K(t)^\dagger ) \, \dd t} \right)\ \middle/\ 
	\left( 1 - \frac{p \efficiency}{2} \left(1 - \frac{\int_0^\infty \left(\ee^{-\Gamma t} \tr(K(t) \rho_M K(t)^\dagger) \right)^2 \dd t}{\efficiency^2}\right) \right)\right. \, ,
\end{equation}
which readily gives \cref{eq:single-photon-fidelity} in the limit of low collection efficiency.

\section{Two-photon scheme general post-herald state}
\label{sec:two-photon-post-herald-derivation}

In this section, we give the general expressions for the atom–atom state resulting from two-photon-heralded protocols. In these protocols, emission from two nodes is combined in a partial Bell-basis measurement setup (see \cref{fig:two-photon-timing}a, \cref{fig:two-photon-geometry}a), such that a detection of a pair of photons in two different detectors heralds the creation of an atom–atom entangled state. Where detectors can easily be made sensitive to the photonic qubit state, for instance through polarising beamsplitters in the case of polarisation encoding, this allows a particular Bell state to be identified with probability one half (in the other half, the state is non-entangled), though the model also applies e.g.~to frequency encoding where a state can only be identified with probability one quarter. Here, as in the main text, we treat the case of polarisation encoding as the paradigmatic example and label the two modes as $H$ and $V$.

As our focus is the role of the motional degree of freedom, we assume that the necessary beamsplitters and detectors are perfect (apart from some amount of loss/limited quantum efficiency, which does not modify the resulting states, only the overall efficiency).
In this case, the post-herald state only involves components where a photon was present from each source; in the wavepackets from \cref{eq:two-photon-wavepacket}, the $\rho_{IM, \textrm{dark}}$ component can be neglected, and we can consider the pure input wavepackets for a pure initial motional state $\ket{\zeta}_{M^s}$ (the $s \in \{A, B\}$ superscript labels the two nodes):
\begin{equation}
  \label{eq:two-photon-state-derivation-wavepacket}
  \icol{\ket{0}}_{I^s} \int_0^\infty \ee^{-\left(\frac{\Gamma}{2} + \ii \icol{\omega_{e0}^s}\right) t} \mcol{K_{sH}}(t) \mcol{\ket{\zeta}_{M^s}} \pcol{\tilde{f_{sH}}^\dagger}(t)\, \dd{t} \pcol{\ket{\vac}} +
  \icol{\ket{1}}_{I^s} \int_0^\infty \ee^{-\left(\frac{\Gamma}{2} + \ii \icol{\omega_{e1}^s}\right) t} \mcol{K_{sV}}(t) \mcol{\ket{\zeta}_{M^s}} \pcol{\tilde{f_{sV}}^\dagger}(t)\, \dd{t} \pcol{\ket{\vac}}.
\end{equation}
Here, as in \cref{sec:two-photon-errors}, we have taken the kick operators $\mcol{K_{si}}(t)$ to already include the effect of the excitation laser kick. We also assume the wavepackets from both nodes to be aligned temporally in the heralding station, and the qubit frequencies to be matched, $\omega_{ei}^A = \omega_{ei}^B$ (here, as in the entire document, we work in the rotating frame w.r.t.~the internal state, so this equality causes the time-dependent phase in the wavepacket to factor out into a global phase).

For a perfect Bell-basis measurement setup, the heralded entangled state is the same for both $H$/$V$ detector combinations where the single-photon detectors on the same, or on opposite sides of the central $50:50$ beamsplitter click. Considering the same-side cases, the sub-normalised state of the internal state and motion conditional on a detection of the photons at times $t_H, t_V \geq 0$ is
\begin{equation}
  \label{eq:two-photon-conditional-proj}
  \rho_{\land \textrm{same}}(t_H, t_V) = \ee^{-\Gamma (t_H + t_V)} \operatorname{proj}\Bigl(\ket{0}_{I^A}K_{AH}(t_H)\ket{\zeta}_{M^A} \otimes \ket{1}_{I^B}K_{BV}(t_V)\ket{\zeta}_{M^B} + \ket{1}_{I^A}K_{AV}(t_V)\ket{\zeta}_{M^A} \otimes \ket{0}_{I^B}K_{BH}(t_H)\ket{\zeta}_{M^B}\Bigr).
\end{equation}
Note that the argument to $\operatorname{proj}(\ldots) = \proj{\ldots}$ is itself in general not normalised, as the kick operators e.g.~contain the collection efficiency. The trace of $\rho_{\land \textrm{same}}(t_H, t_V)$ gives the probability density for observing herald photons at the given times, and the normalised state after having detected such a pair of photons is $\rho_{| \textrm{same}}(t_H, t_V) = \rho_{\land \textrm{same}}(t_H, t_V) / \tr(\rho_{\land \textrm{same}}(t_H, t_V))$. The state for detector clicks on opposite sides, $\rho_{\land \textrm{opp}}(t_H, t_V)$ is very similar, just with a minus sign in the superposition in \cref{eq:two-photon-conditional-proj}; we will discuss the $\rho_{\land \textrm{same}}(t_H, t_V)$ case for brevity.

As in \cref{sec:single-photon-post-herald-derivation}, \cref{eq:two-photon-conditional-proj} is easily generalised to arbitrary initial motional states through their spectral decomposition. After discarding the motional states, the post-herald state in the computational basis becomes
\begin{multline}
  \label{eq:two-photon-conditional-motiontrace}
  \rho_{\land \textrm{same}}(t_H, t_V) = \ee^{-\Gamma (t_H + t_V)} \cdot\\
  \begin{pmatrix}
    0 & 0 & 0 & 0 \\
    0 & \tr(K_{AH}(t_H)\,\rho_{M^A} K_{AH}(t_H)^\dagger) \tr(K_{BV}(t_V)\,\rho_{M^B} K_{BV}(t_V)^\dagger) & \tr(K_{AH}(t_H)\,\rho_{M^A} K_{AV}(t_V)^\dagger) \tr(K_{BV}(t_V)\,\rho_{M^B} K_{BH}(t_H)^\dagger) & 0 \\
    0 & \tr(K_{AV}(t_V)\,\rho_{M^A} K_{AH}(t_H)^\dagger) \tr(K_{BH}(t_H)\,\rho_{M^B} K_{BV}(t_V)^\dagger)  & \tr(K_{AV}(t_V)\,\rho_{M^A} K_{AV}(t_V)^\dagger) \tr(K_{BH}(t_H)\,\rho_{M^B} K_{BH}(t_H)^\dagger)  & 0 \\
    0 & 0 & 0 & 0 \\
  \end{pmatrix}.
\end{multline}
Averaging over all possible detection times gives $\rho_{\land \textrm{same,all}} = \int_{0}^{\infty}\int_{0}^{\infty} \rho_{\land \textrm{same}}(t_H, t_V)\, \dd{t_V}\, \dd{t_H}$ and $\rho_{| \textrm{same,all}} = \rho_{\land \textrm{same,all}} / \tr(\rho_{\land \textrm{same,all}})$.

The fidelity to the closest maximally entangled state of the normalised density matrix is $F(\rho_{| \textrm{same,all}}) = \frac{1}{2}\left(1 + |C|\right)$, where
\begin{multline}
   C = 2 \int_{0}^{\infty}\int_{0}^{\infty} \ee^{-\Gamma (t_H + t_V)} \tr(K_{AH}(t_H) \rho_{M^A} K_{AV}(t_V)^\dagger) \tr(K_{BV}(t_V) \rho_{M^B} K_{BH}(t_H)^\dagger)\, \dd{t_V}\, \dd{t_H}\ \Bigg/ \\
  \Bigg(
    \int_{0}^{\infty}\ee^{-\Gamma t_H} \tr(K_{AH}(t_H) \rho_{M^A} K_{AH}(t_H)^\dagger) \, \dd{t_H}
    \int_{0}^{\infty}\ee^{-\Gamma t_V} \tr(K_{BV}(t_V) \rho_{M^B} K_{BV}(t_V)^\dagger) \, \dd{t_V}\ + \\
    \int_{0}^{\infty}\ee^{-\Gamma t_V} \tr(K_{AV}(t_V) \rho_{M^A} K_{AV}(t_V)^\dagger) \, \dd{t_V}
    \int_{0}^{\infty}\ee^{-\Gamma t_H} \tr(K_{BH}(t_H) \rho_{M^B} K_{BH}(t_H)^\dagger) \, \dd{t_H}
  \Bigg).
\end{multline}
If the two systems and initial states are identical, this simplifies to
\begin{equation}
  C = \frac{\int_{0}^{\infty}\int_{0}^{\infty} \ee^{-\Gamma (t_H + t_V)} \absval{\tr(K_{H}(t_H) \rho_{M} K_{V}(t_V)^\dagger)}^2 \dd{t_V}\, \dd{t_H}}{\int_{0}^{\infty}\ee^{-\Gamma t_H} \tr(K_{H}(t_H) \rho_{M} K_{H}(t_H)^\dagger)\, \dd{t_H}
  \int_{0}^{\infty}\ee^{-\Gamma t_V} \tr(K_{V}(t_V) \rho_{M} K_{V}(t_V)^\dagger)\, \dd{t_V}},
\end{equation}
and $C$ is then real and non-negative, giving \cref{eq:two-photon-symmetric-fidelity} from the main text.

Clearly, the post-herald states and fidelities for constrained detection times can be obtained by adjusting the integral bounds accordingly. (In practice, this might be implemented – apart from some nominal cutoff many time-constants after excitation – to reduce the impact of non-idealities such as dark counts).

If the motional degrees of freedom are not immediately ignored, but an operation is applied by either node to its atom's motional state, this is straightforwardly multiplied from the left to the non-normalised pure state in \cref{eq:two-photon-conditional-proj}, effectively modifying the kick operators. As discussed in \cref{sec:two-photon-errors}, by making this motional operation conditional on the internal state and adjusting it in real-time according to the photon detection times $t_H$ and $t_V$, one can attempt to disentangle spin and motion as well as possible given the constraints on the realisable correction operators. If the single-photon detectors do not have a fine time resolution compared to the relevant time scale (given by the motional frequencies or the mismatch in qubit frequencies, if any), this limits the accuracy with which $t_H$ and $t_V$ can be known, and hence the correction can be applied.

Note that \cref{eq:two-photon-state-derivation-wavepacket} is not the most general wavepacket possible, as we have taken the two decay channels to correspond to entirely orthogonal output modes. This gives rise to the simple forms of \cref{eq:two-photon-conditional-proj,eq:two-photon-conditional-motiontrace} only in the $\{\ket{01}, \ket{10}\}$ subspace. If this is not a good approximation, either because of intrinsic limitations in a particular emission/collection setup, or because some symmetry properties of the system might no longer hold in the presence of motion (such as discussed in \cref{sec:high-na-orthogonal-errors}), the above can be easily adapted by expanding $\ket{0}_{I^A} K_{AH}(t_H) \mapsto (\ket{0}_{I^A} K_{A0H}(t_H) + \ket{1}_{I^A} K_{A1H}(t_H))$ etc.~in \cref{eq:two-photon-conditional-proj}.

If necessary, the model for motional effects discussed here can straightforwardly be extended by non-ideal optical elements, temporal misalignment, a mismatch in qubit frequencies, detectors with dark counts, or similar imperfections; see \cite[ch.~4.4]{nadlingerDeviceIndependentQuantumKey2021}.

\section{Geometric effects in high-NA collection from crossed dipoles}
\label{sec:high-na-orthogonal-errors}
\label{sec:longitudinal-polarisation-symmetry-loss}

\begin{figure}
  \centering
  \includegraphics{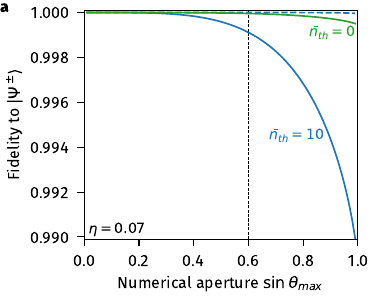}%
  \hspace{12pt}%
  \includegraphics{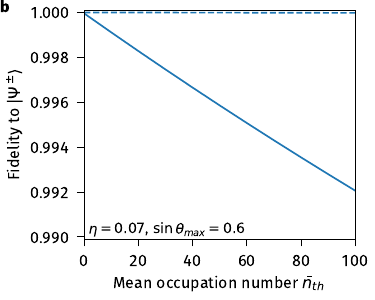}
  \caption{Loss in fidelity from geometric effects in two-photon entanglement from $\pi/\sigma$ decays (ignoring temporal effects, i.e.~$\Gamma \rightarrow \infty$), with $\eta = 0.07$ typical of trapped-ion experiments. \figpart{(a)} As the \na{} is increased, the error from $\sigma$ emission giving rise to polarisation aligned with the quantisation axis (solid lines), while the error just from the mismatch of dipole emission patterns remains negligible (\num{3e-5} at \na{} 1.0 and $\nbar_{th} = 10$; dashed line). \figpart{(b)} For fixed \na{} 0.6, dependence of the fidelity on the initial motional state mean occupation number $\nbar_{th}$.}
  \label{fig:high-na-pi-sigma-errors}
\end{figure}

In the discussion of two-photon protocols in the main text, we have focussed on errors due to the spread in detection times (\cref{sec:two-photon-timing-errors}), and considered errors from the collection geometry only when the two decay channels are collected from different directions (\cref{sec:two-photon-geometry-errors}). However, when photons from decays corresponding to two orthogonal dipoles are collected from large solid angles (high \na{}s), there are error mechanisms that affect even collection along the same optical path into e.g.~the same single-mode fibre.

Two such mechanisms commonly exist. First, consider the case where the emitting dipoles are $\vvn{d}_{e0} = \vvn{x}$ and $\vvn{d}_{e1} = \vvn{y}$ and collection is along $\vvn{z}$. The collection weights for the two decays differ away from the optical axis due to the asymmetry of the dipole radiation patterns (e.g.~$w_{0H}(\vvn{k})$ will fall off more quickly in the $x$ than the $y$ direction, as the radiation pattern from an $x$-oriented dipole has a node along $\vvn{x}$). Secondly, in the case of ion–photon entanglement generation between $J=1/2 \leftrightarrow J'=1/2$ levels perpendicular to the magnetic field direction~\cite{Blinov2004,stephensonHighRateHighFidelityEntanglement2020}, technically advantageous due to its simplicity, the two decay channels correspond to $\sigma^\pm$ and $\pi$ dipole transitions with $\vvn{d}_{e0} = -\left(\vvn{x} \pm \ii \vvn{z}\right) / \sqrt{3}$ and $\vvn{d}_{e1} = \vvn{y} / \sqrt{3}$ (magnetic field along $\vvn{y}$). When the emitter is exactly placed on the optical axis, emission from the longitudinal $\vvn{z}$ dipole that is a component of the circular $\sigma$ dipole does not couple into a azimuthally
symmetric fibre mode~\cite{stephensonHighRateHighFidelityEntanglement2020,Kim2011}, enabling high-fidelity entanglement generation. In the presence of motion, this cancellation is no longer exact, leading to errors.

Our framework naturally covers arbitrary collection geometries in the full vector-optics picture through appropriate calculation of the $w_{i\sigma}(\vvn{k})$ weights. Instead of \cref{eq:two-photon-state-derivation-wavepacket} with the diagonal decay channel/polarisation mapping, the full wavepackets from \cref{eq:collected-state} with the sum over polarisations need to be used, but the rest of the calculation proceeds analogously (giving more non-zero terms in the equivalent of \cref{eq:two-photon-conditional-motiontrace}). A numerical evaluation for parameters typical of trapped-ion experiments ($\eta = 0.07$), coupled into an optimally-sized Gaussian mode, is shown in \cref{fig:high-na-pi-sigma-errors}. The first mechanism discussed (mismatch of the transversal dipole emission patterns) remains negligible, reaching only an error of \num{3e-5} in the limit of \na{} 1.0 at $\nbar_{th} = 10$. Similar to the effect discussed in \cref{sec:two-photon-geometry-errors}, the fact that the superposition of displacement operators is symmetric around the plane perpendicular to the dipole suppresses the associated errors; they could be feasibly reduced somewhat further by the application of state-dependent squeezing (and higher-order terms). The otherwise-suppressed longitudinal dipole coupling dominates the errors and is non-negligible at moderate temperatures; it reaches about $\SI{0.1}{\percent}$ at \na{} 0.6, $\eta = 0.07$ and $\nbar_{th} = 10$.

Thus, for typical trapped-ion experiments operating close to the Doppler limit, these effects are not entirely negligible, but do not currently dominate error budgets. For more weakly trapped emitters, or at higher temperatures, these effects deserve scrutiny, however; working with e.g.~pure $\sigma^\pm$ emission imaged along the magnetic field, frequency encoding, etc.~where there is no longitudinal dipole component to reject, may yield better results. In typical atomic structures, if two such $\sigma^\pm$ decays are possible, a third $\pi$ decay channel will also be present that must then be rejected via symmetry, suffering from the same thermal dependence if such decays are not explicitly selected out in a post-herald check (our calculations allow for an arbitrary number of lower states, and can thus be straightforwardly extended to such scenarios). Excitation into a superposition of levels with only one allowed decay channel each (e.g. $\sigma^+$ and $\sigma^-$ on a $J=1/2 \leftrightarrow J'=3/2$ transition), or a time-bin encoding scheme making repeated use of the same transition, may avoid the polarisation mixing effects through filtering, though the timing (phase-modulation) effects of large thermal states will still lead to errors.

\twocolumngrid
\bibliography{spontaneous-emission-recoil}

\end{document}